\pdfoutput=1
\documentclass[9pt,sigconf]{acmart}
\usepackage{algorithm}
\usepackage[noend]{algpseudocode}
\usepackage{paralist}
\usepackage{epsfig} 
\usepackage{soul}
\usepackage{mathtools}
\usepackage{amsmath} 
\usepackage{amssymb}  
\usepackage{graphicx}
\usepackage[export]{adjustbox}
\usepackage{geometry}
\usepackage{subcaption}
\usepackage{listings,multicol}
\usepackage{lipsum}
\usepackage{setspace}
\usepackage{comment}
\usepackage{multirow}
\usepackage{booktabs}
\usepackage{tcolorbox}
\usepackage{color}
\usepackage{color, colortbl}
\usepackage{enumitem}
\usepackage{epstopdf}
\usepackage{siunitx}
\usepackage[font={small}]{caption}
\usepackage{float}

\newcommand{\q}[1]{\lq #1\rq}
\usepackage{wrapfig,lipsum,booktabs}

\DeclarePairedDelimiter{\ceil}{\lceil}{\rceil}
\DeclarePairedDelimiter{\floor}{\lfloor}{\rfloor}
\DeclarePairedDelimiter{\norm}{\Vert}{\Vert}

\DeclarePairedDelimiter{\tupleangle}{\langle}{\rangle}
\usepackage{setspace}

\definecolor{listinggray}{gray}{0.9}
\definecolor{lbcolor}{rgb}{0.9,0.9,0.9}
\newtheorem{Definition}{Definition}

\title{Skip to Secure: Securing Cyber-physical Control Loops with Intentionally Skipped Executions} 
\author{Sunandan Adhikary\footnotemark[1], Ipsita Koley\footnotemark[1], Sumana Ghosh\footnotemark[1], Saurav Kumar Ghosh\footnotemark[1], Soumyajit Dey\footnotemark[1], Debdeep Mukhopadhyay\footnotemark[1]\\
	*{\it Indian Institute of Technology, Kharagpur}\\ {\tt\small \{sunandana,ipsitakoley,soumyajit,debdeep\}@iitkgp.ac.in,\{sumanaghosh,saurav.kumar.ghosh \}@cse.iitkgp.ernet.in}   
}

\setcopyright{none}
\renewcommand\footnotetextcopyrightpermission[1]{}
\acmYear{2020}

\begin{document}

\begin{abstract} 
We consider the problem of provably securing a given control loop implementation in the presence of adversarial interventions on data exchange between plant and controller. Such interventions can be thwarted using continuously operating monitoring systems and also cryptographic techniques, both of which consume network and computational resources. We provide a principled approach for intentional skipping of control loop executions which may qualify as a useful control theoretic countermeasure against stealthy attacks which violate message integrity and authenticity. As is evident from our experiments, such a control theoretic counter-measure helps in lowering the cryptographic security measure overhead and resulting resource consumption in Control Area Network (CAN) based automotive CPS without compromising performance and safety. 
\end{abstract}


\maketitle
\section{Introduction} 
\label{sec:intro}
The proliferation of network connectivity has increased the application domain for cyber-physical systems (CPS) in today's connected world. However,  increased connectivity manifests security vulnerability in terms of increased number of possible attack surfaces for such systems. Recent results have established that network based {\em Man-in-the-Middle} type attacks, like false data injection on cyber-physical control systems are quite capable of disturbing closed loop stability as well as degrading the control performance of such systems \cite{teixeira2015secure1}. 
In such an attack, an adversary injects false data in the communication medium between the plant and the controller with the intention of driving the system to an unsafe state by changing the set point of the system.

\par \textbf{State-of-the-art detection systems:} In order to detect such attacks, the most common control theoretic countermeasures put in place are threshold based anomaly detectors that generate an alarm if the estimation error crosses the threshold over a single or multiple control loop iterations. Though such control theoretic primitives can limit the attacks,
it has been observed that stealthy attacks are possible even in the presence of such state estimation based lightweight  control theoretic intrusion detectors  
\cite{teixeira2015secure2}. Hence, the standard way to secure a system against such attack is the use of security 
primitives like Message Authentication Codes (MACs) \cite{munir2018design}, Message
Encryption \cite{munir2018design}, Physically Unclonable Functions (PUFs) \cite{ghosh2018performance} etc. 
Some recent efforts also focus on learning based \cite{kreimel2017anomaly, vatanparvar2019self} intrusion detection mechanisms.
However, the options for implementing such security primitives in CPS is often limited by the available compute resources in the on-board platforms. Hence, there have been proposals~\cite{jovanov2017sporadic} for \textit{sporadic usage} of such \textit{Intrusion Detection Systems} (IDS) for securing messages exchanged between software-based controllers and physical plants.

\par
\begin{wrapfigure}{r}{0.35\columnwidth}
	\centering
	\includegraphics[scale=0.7,clip]{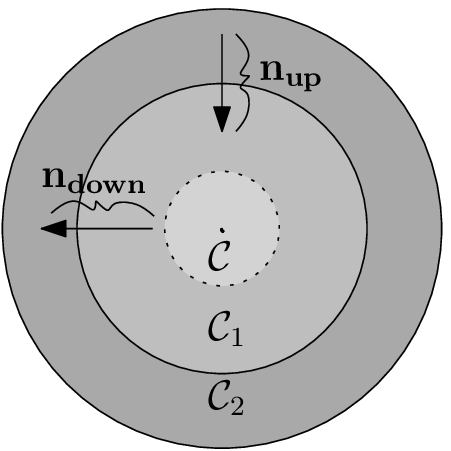}
	\caption{\footnotesize Sporadic IDS}
	\label{fig:ids}
\end{wrapfigure}

 \textbf{CPS operation under Sporadic IDS:}  A sporadic IDS can be specified by a pair $(n_{up}$,  $n_{down})$ such that  the IDS is active for $n_{up}$ consecutive control samples and inactive for $n_{down}$ consecutive control iterations, and this behavior repeats in a cycle. As shown in  Fig.\ref{fig:ids}, let for a control system, there exists an {\em initial region} $\mathcal{C}$  which is composed of the initial range of plant state values.
Starting from $\mathcal{C}$, consider that the preferable operating region for the system is given by an {\em inner safety region} $\mathcal{C}_1 (\mathcal{C}\subseteq\mathcal{C}_1)$ in the absence of any external attacks. The  safety guarantee offered by a sporadic IDS is based on the existence of an {\em outer safety region} $\mathcal{C}_2$ ($\mathcal{C}_1\subset \mathcal{C}_2$) which meets the safety requirements of the system, but may not be a preferable operating region for unsatisfactory control performance. The IDS parameters, $n_{up}$, $n_{down}$ can be formally  defined as,
{\small
\begin{align*}
    x[k]\in \mathcal{C}_1 \implies & \forall i\leq n_{down},\, x[k+i] \in \mathcal{C}_2 \mbox { when IDS is off}\\
    x[k]\in \mathcal{C}_2 \implies & \forall i\geq n_{up},\, x[k+i] \in \mathcal{C}_1 \mbox { when IDS is on}
\end{align*}
}
where $x[k]$ denotes the plant state at any time instant $k$.
When an IDS is not available for $n_{down}$ consecutive control iterations, stealthy attacks (similar to \cite{jovanov2017sporadic}) which the control system is hoodwinked to think as environmental noise are possible. The period $n_{down}$ should be small enough to ensure that starting from $\in \mathcal{C}_1$, such attacks should not steer the system outside $\mathcal{C}_2$. When the IDS is active for $n_{up}$ consecutive control iterations, no false data injection attack is possible. The period $n_{up}$ needs to be large enough to ensure that the system is brought inside $\mathcal{C}_1$ starting from anywhere $\in \mathcal{C}_2$.  This ensures that the system duly recovers from the effect of false data injected during the period when IDS was inactive thus nullifying attacker's  efforts. 

\par \textbf{Attack resilience of a system:} Attack resilience of an IDS enabled CPS is measured by the value of $n_{down}/n_{up}$.  
Let $d_{min}$ be the \textit{minimum attack-length}, i.e., the  minimum number of consecutive control samples required by an attacker to drive the system out of $\mathcal{C}_2$ (starting $\in \mathcal{C}_1$) while remaining undetected (thus defining a {\em minimum effort} successful attack). We can bound the down-time $n_{down}$ of an IDS as $n_{down}<d_{min}$. This allows us to set a maximum down time of $n_{down}=d_{min}-1$ in order to stop the attacker before being successful. Thus, increasing $d_{min}$ with suitable choice of CPS parameters in-turn increases the attack  resilience (i.e., $n_{down}/n_{up}$) of the system.  Furthermore, the increment in $n_{down}$ proportionally reduces the computational and communication requirement of the IDS. 
Considering resource-constrained CPS implementation platforms, it is always desirable to go for lightweight provably secure IDS schemes by maximizing  down-time (i.e., $n_{down}$) without sacrificing safety and and performance in the presence of stealthy attacks, which is the focus of this work. 

\par \textbf{Motivation and Problem statement:} Computing control law over falsified sensor  measurements can actually drive a CPS towards an unsafe state. Hence, in order to  minimize the  effect of false data injection in sensor measurements, it may be useful to skip the control law computation in some carefully chosen sampling instants while ensuring that such occasional {\em execution skips} do not hamper the desired control performance. 
The system does not get affected by  malicious data injected by the attacker into the communication channel at  sampling instants when the control executions are skipped. So, even if the attacker is aware of the positions of skipped executions, it has to try longer to make the system unsafe by fault data injection. When  the  system is running following some carefully chosen \textit{control skipping pattern} unknown to attacker, 
the  attacker may potentially require longer periods of attack efforts to guess the skip positions and efficiently inject faulty  data into the system in order to succeed. 
 In the present work, we motivate employing execution skips as a \textit{secure control mechanism}. Our  proposed framework considers a CPS specification and automatically synthesizes \textit{control skipping patterns}  which maximize the attack resilience  without compromising the desired control performance of the system. The synthesis process also provides us an IDS activation schedule with minimized computational cost as a by-product.

\par \textbf{Proposed approach and Contributions:}
The above mentioned goals require setting up and solving a non-linear multi-objective optimization problem. The problem is non-trivial  since we want to 1) maximize attack  resilience, while also retaining 2) the control performance as much as possible. Note that both these objectives are highly dependent on the positions of execution skips in the control schedule and they do not follow a monotonic relationship. The pattern exhibiting best control performance may lack in attack resilience. 
Also, the dependence of control performance on the  skipping pattern of a control schedule is nonlinear  \cite{ghosh2017structured}. Hence, formulating a single step optimization framework for maximizing both control performance and attack resilience of a CPS is not a scalable approach. For this reason, we propose a two-step optimization framework. In the first step, we synthesize a set of control skipping patterns that are ranked based on their control performance. In the next step, we analyze  the attack resilience of these patterns using  Satisfiability Modulo Theory (SMT) based techniques. Higher attack resilience in-turn guarantees less usage of IDS along with the underlying computing and communication platforms. In summary, our contributions can be listed as follows.
    \par\noindent \textbf{(1)} We present the first work  that motivates the use of intentional \textit{execution skips} as a control-theoretic security measure.
    \par\noindent \textbf{(2)} In order to formally analyze the robustness of this measure, we build an SMT based algorithmic framework for synthesizing  successful but stealthy false data injection attack vectors.
    \par\noindent \textbf{(3)} We leverage this framework for designing  sporadic IDS with increased down-time (or more attack resilience) when compared with existing sporadic IDS schemes used with period control implementations  (i.e., without execution skips) \cite{jovanov2017sporadic}.
    \par\noindent \textbf{(4)} 
    Since the pattern search space is exponential in pattern length, we develop a pruning mechanism 
     for classifying control skipping patterns based on their offered performance. This step is instrumental in rendering our method scalable for sporadic IDS design.
    \par\noindent \textbf{(5)} We establish the usefulness of our approach by considering automotive system examples where  sporadic IDS solutions generated by our tool set provided performance and security guarantees similar to previously reported schemes while consuming less communication bandwidth and computational resources.
\section{Model Description}
\label{sec:model_des}
This section briefly describes the model of the plant and controller, followed by mathematical description of CPS under attack.
\subsection{Control System Modeling}
\label{subsec:system_model} A physical plant can be represented as a linear discrete-time  invariant system (LTI) having the dynamical equations given as follows. 
{\small
\begin{align}
\label{eq:state_space}
x[k+1] &= Ax[k] + Bu[k],\ \ \
y[k+1] = Cx[k+1]\\\nonumber
\hat{x}[k+1] &= A\hat{x}[k]+Bu[k]+L(y[k]-C\hat{x}[k]),
u[k+1] = K\hat{x}[k+1]
\end{align}
}
Here $x[k]$ is the value of state variable at $k$-th iteration, which is being controlled by  control input $u[k]$ calculated by the controller  based on the estimated state $\hat{x}[k]$. In this work, we consider Kalman Filter \cite{kalman1960new} for state estimation and Linear Quadratic Regulator (LQR) based optimal control technique for calculating the control input. The control input is received by actuators in plant side and control action can not be exerted beyond the actuator saturation limit. 
In Eq.~\ref{eq:state_space}, the estimated state is calculated using the Kalman Gain, $L$ and output measurement $y[k]$. 
Plant outputs are  sampled by sensors and transmitted provided they are within supported sensing ranges. The matrices $A,B,C,D$ are system  matrices and constant in nature. For a plant-control loop $(P, K)$ with $K$ as the state feedback gain, we define $X[k]=[x^{\textsf{T}}[k]~ \hat{x}^{\textsf{T}}[k]~ u[k]^{\textsf{T}}]^{\textsf{T}}$ as state vector for the augmented system comprising both the plant and estimator states along with control inputs. 
The augmented system helps analyze the effect of execution skips on the closed loop. The dynamical equation for the augmented system is given by, $X[k+1]= A_{1}~X[k]$, where 
$A_{1}=\left[{\small \begin{array}{ccc}A& 0& B\\LC& A-LC-BK& 0\\KLC & KA-KLC-KBK& 0\end{array}}\right]$. If the execution of the controller is intentionally skipped inside a sampling interval $[k, k+1)$, 
no new control update is communicated to the plant and state estimation unit in that sampling instance but sensor update is received. Therefore,  the plant state is updated using the last communicated control input from previous iteration i.e., $u[k+1] = u[k]$ and state space equations change as follows.
{\small
\begin{align}\nonumber
 x[k + 1] =& A~x[k]+B~u[k],\ \ u[k+1]=u[k]\\
\hat{x}[k+1]
=&LC~x[k]+(A-LC)~\hat{x}[k]+Bu[k]
\label{eq:loop_matrix_drop}
\end{align}
}
\noindent Following Eq.\eqref{eq:loop_matrix_drop}, during control skips the augmented system progresses with ${\small A_{0}=\left[{\small \begin{array}{ccc}A& 0& B\\LC& A-LC-BK& 0\\0 & 0& I\end{array}}\right]}$ instead of $A_1$. Next, we define the notion of \textit{control skipping pattern} as follows. 
\begin{Definition}
\textbf{Control Skipping Pattern :} An $l$-length control skipping pattern for a given control loop $(P,K)$, is an $l$ length sequence $\rho\in \{0, 1 \}^l$ such that it can be used to define an infinite length control schedule $\pi=\rho^\omega$, repeating with period $l$, i.e., $\pi[k]=\pi[k+l] = \rho[k\%l], \forall k\in \mathbb{Z}^+$. \hfill$\Box$
\label{def:loop_skipping_pat}
\end{Definition}
\noindent The evolution of the closed loop system according to a control skipping pattern can be exemplified as: for $\rho= 110010$, we have, 
{\small \[ X[6] =A_{1} X[5]=A_{1} A_{1} A_{0} X[3] = \ldots = A_{1} A_{1} A_{0} A_{0} A_{1} A_{0} X[0]. \]}

\subsection{Control Design and Performance Metrics}
\label{subsubsec:stability}

The control design metric represents the control objective while designing the controller. One such design metric that we use in this work is \textit{settling time}. It is the time needed by the system output to fall and stay around the reference value (e.g., within \SI{2}{\percent} error band). Hence, the controller has to be designed in such a way that given settling time requirement is always met. On the other side, the \emph{ control performance} is the measure of quality of control (QoC), i.e., how efficiently the design requirement is met. In this work we consider LQR-based controller design technique. So we use \textit{LQR cost function} $J$ as the performance metric  given by,~
{\small
$\label{eq:lqr_cost}
J= \sum_{k=0}^\infty (x^\mathsf{T}[k]Qx[k] + u^\mathsf{T}[k]Ru[k]),
$}
~\cite{astrom97}, with $Q \succcurlyeq 0$ and $R \succ 0$ being symmetric weighing matrices capturing the relative importance that the control designer can give to the state deviation and control effort respectively. Lower the LQR cost better the performance. 

\begin{wrapfigure}{r}{0.4\columnwidth}
	\centering
	\includegraphics[width=1.05\linewidth,clip]{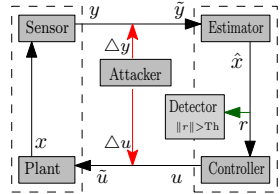}
	\caption{CPS attack model}
	\label{fig:secureControl}
	\vspace{2pt}
\end{wrapfigure}

A significant amount of work exists in the literature addressing the issue of control design and performance in the presence of execution skips  \cite{zhang01,ghosh2017structured,soudbakhsh_13}. 
Given the settling time requirement, $T_s$, we follow Theorem 4.1 of \cite{ghosh2017structured} to calculate the \textit{minimum execution rate}, $r_{min}$, from $T_s$. This essentially means, to maintain $T_s$, the controller has to be executed at least $\lceil l\times r_{min}\rceil$ times in  $l$-length consecutive control samples, i.e., in an $l$-length control skipping pattern, $\rho$, there has to be at least $\lceil l \times r_{min}\rceil$ number of \lq $1$\rq s. On the other hand, control performance varies with relative  positions of the execution skips in a pattern (i.e., distribution of \lq $0$\rq s over $\rho$)~\cite{jia2007graceful}.


\subsection{Attack Modeling}
\label{subsec:attack_model}

A schematic of a cyber-physical system under stealthy false data injection attacks is given in Fig.~\ref{fig:secureControl}. We consider a stealthy attack scenario where the communication network has been compromised and an adversary can (i) provide false sensor measurements to the controller, denoted by $\tilde{y}[k]=y[k]+\triangle y[k]$ and (ii) tamper with the control input resulting in  $\tilde{u}[k]=u[k]+\triangle u[k]$ received by the actuators. Here, $\triangle y[k]$ and $\triangle u[k]$ are the amount of measurement and actuation errors respectively, induced by the attacker at the $k$-th iteration, and we express this with an attack vector, $\mathcal{A}[k]=[\triangle{u}^{\textsf{T}}[k]~\triangle{y}^{\textsf{T}}[k]]^{\textsf{T}}$.  Under these circumstances, the estimator estimates corrupted $\hat{x}[k+1]$ (i.e., $\tilde{\hat{x}}[k+1]$) to minimize the residue $r[k]=\tilde{y}[k]-C\hat{x}[k]$ (i.e., the difference between the measurement received and the estimated measurement). Due to such a compromised control sample, the plant states are polluted by the attacker-induced errors. As a result, the manipulated states $\tilde{x}[k]$ are driven towards an unsafe region (outside of $\mathcal{C}_2$). We can formalize the state progression in attacked situation using our augmented system with manipulated state vector, {\small $\tilde{X}[k+1]=A_1~\tilde{X}[k]+B_1~\mathcal{A}[k]$} where, {\small $B_1^{\textsf{T}}=\begin{bmatrix}\scriptsize 0 &L^{\textsf{T}} &L^{\textsf{T}}K^{\textsf{T}}\\0 &0 &I \end{bmatrix}$}. In presence of execution skip, $B_1^{\textsf{T}}$ can be replaced with $B_0^{\textsf{T}}=\begin{bmatrix}0 &L^{\textsf{T}} &0 \\\nonumber 0 &0 &0 \end{bmatrix},$ causing minimized perturbations during skipped executions. 
Note that to the plant and controller these false data injections may get disguised as process and measurement noises. 
Following existing techniques for {\em physics based attack} detection \cite{giraldo2018survey}, we assume the following system protection and attack model.
\par\textbf{(1)} In our protection system model, the threshold-based intrusion detector flags an attack whenever the residue $r[k]$ surpasses the detector threshold given by some constant $Th$, i.e., $\norm{r[k]}> Th$, which in turn limits the attacker's effort of manipulation ($\norm{.}$ denotes vector 2-norm). We can also consider the system to be fitted with popularly used $\chi^2$ based attack detectors  since detection criteria in such probabilistic detectors can as well be interpreted as  non-probabilistic threshold-based detection techniques \cite{jovanov2017sporadic}.

\par \textbf{(2)} The attacker has full knowledge of the system dynamics and threshold-based detectors present in the system. The attacker can observe the system closely and choose proper false data irrespective of knowing the control skipping pattern. The system supported sensor range and actuator saturation limit impose a bound on attacker's stealthy efforts.
\par\textbf{(3)} The goal of the attacker is to alter the  operating point of the system thereby driving it to an unsafe state $x \notin {\mathcal{C}_2}$ in the least amount of time possible while remaining stealthy. An attack vector of length $d$ can be defined as {\small $\mathcal{A}_d=\mathcal{A}[1:d]=\begin{bmatrix}
\triangle u_1 & \cdots & \triangle u_d\\ 
\triangle y_1 & \cdots & \triangle y_d
\end{bmatrix}$. The attack vector $\mathcal{A}_d$} launched on a protected control system executing its $k$-th iteration is deemed
\textbf{1)} {\em stealthy} if $\Vert r[i]\Vert\leq Th$ for all $i\in[k+1,k+d+n_{up}]$ where $n_{up}$ is the up-time of the IDS, and 
\textbf{2)} {\em successful} if $\exists j\in[k+1,k+d+n_{up}]$ such that $x[j]\notin \mathcal{C}_2$, i.e., it violates the safety criterion of the system.
Note that we define the {\em stealthiness} and {\em success} of an attack of length $d$ over a window of $d+n_{up}$ control samples, because an attack of $d$-iterations can drive the system to an unsafe state even after the attack is over. So, we check the safety criteria for a period equal to the  attack duration $d$ followed by the time $n_{up}$  between the attacker's two consecutive attempts. This setting works because of our additional constraint that during IDS operation for period $n_{up}$ we ensure that the system will converge back inside $\mathcal{C}_1$. 
\section{A Motivating Example}
\label{sec:motivating_example}
We consider a \textit{trajectory tracking control} (TTC) example to demonstrate the advantage of using control skipping pattern in improving the attack resilience of the system. TTC system regulates deviation (denoted by $D$) of a vehicle from a given trajectory and deviation (denoted by $V$) from a reference velocity by applying proper amount of acceleration as control input. To cope up with the space we refer to Tab.~\ref{tab:sysSpecs} for the system matrices and initial safety regions. Following \cite{ghosh2017structured}, the settling time criterion of $\SI{5}{\second}$ allows maximum $50\%$ execution skips, i.e., $r_{min}=0.5$ for this system. The  protection system considered in place is a threshold-based anomaly detector having $Th=2$. The attacker model is as described in  Sec.~\ref{subsec:attack_model}. 

\begin{wrapfigure}{l}{0.6\columnwidth}
 	\begin{subfigure}{0.66\columnwidth}
 	\centering
  	\includegraphics[width=1.05\linewidth,clip,scale=1.5]{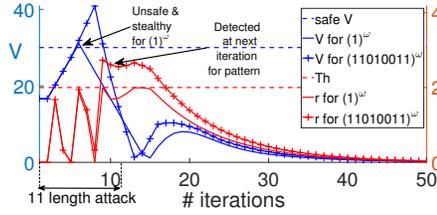}
 \caption{\footnotesize Attack vector for periodic not stealthy on $11010011$}
 \label{fig:attackOnPattaern}
 \end{subfigure}

 \begin{subfigure}{0.66\columnwidth}
 \centering
  	\includegraphics[width=1.05\linewidth,clip]{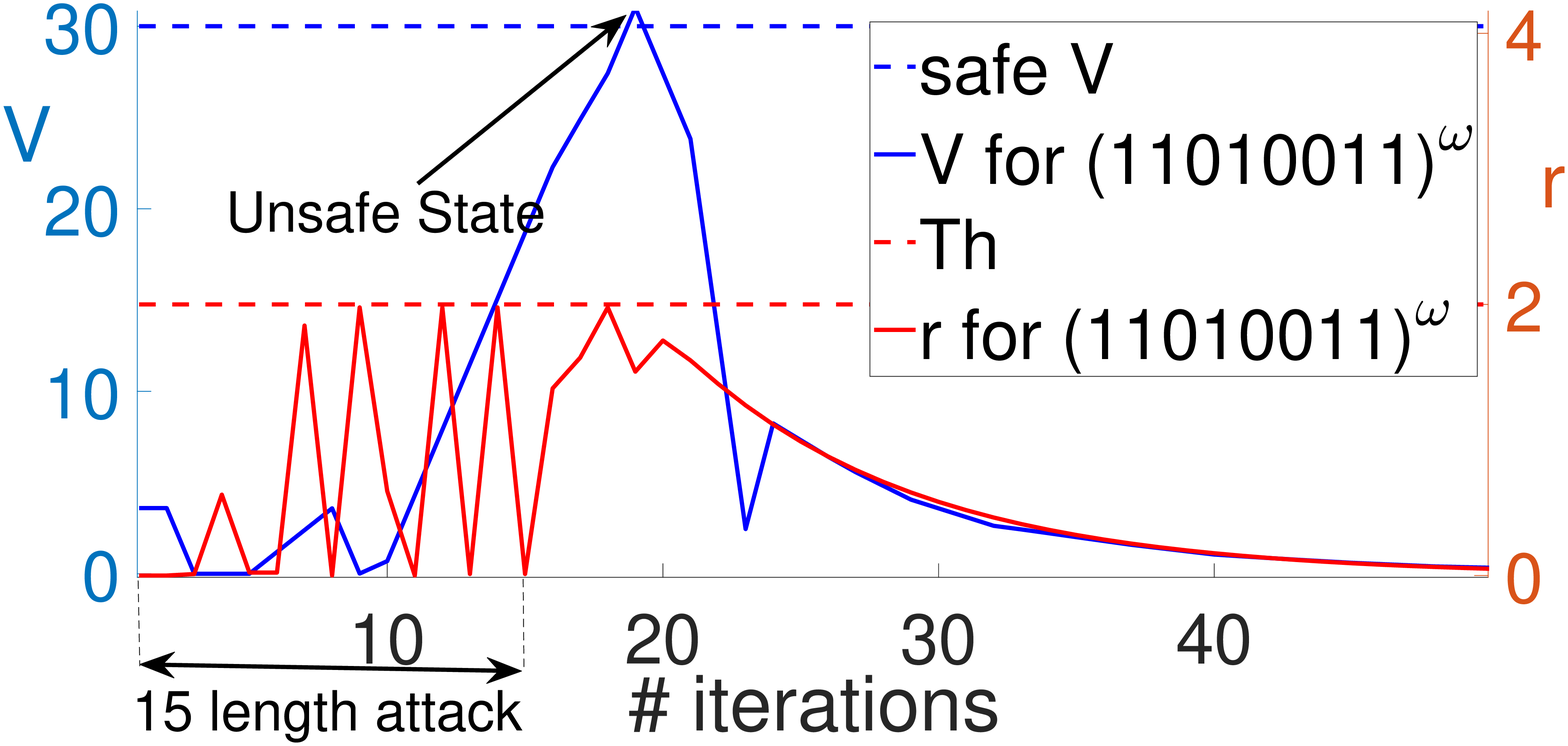}
 \caption{\footnotesize Stealthy and successful attack for $11010011$ with more $n_{down}$} 
 \label{fig:attackOnPattaern1}
 \end{subfigure}
\vspace{-2pt}
 \caption{\footnotesize Plotting $V$ (in blue)  in left y-axis and residue $r$ (in red) in right y-axis (in corresponding scales) to demonstrate the effect of stealthy attack on TTC with and without pattern-based execution. $V$ crossing the blue dashed line (safety boundary of $V$) leads to violation of safety and $r$ crossing the red dashed line ($Th$) indicates attack is detected.}
 \vspace{1pt}
 \label{fig:attack}
\end{wrapfigure}

In Fig.~\ref{fig:attackOnPattaern}, we consider two possible control schedule scenarios. With the periodic pattern $1^\omega$, there exists an attack vector of length 11 for which the system becomes unsafe at the \textit{6-th iteration}. However, this attack vector is stealthy as the residue is never higher than the threshold. The reason that the attack length need to be much larger than the point of safety violation is because, suddenly stopping the attack after the 6-th iteration will lead to large residue and thereby detection. Hence the attack needs to gradually decrease without drastic modification in system dynamics. In fact, it can be checked that for this system, 11 is the {\em minimum attack length} ($d_{min}$), i.e. there does not exist any attack vector of length $< 11$ which is stealthy but successful.
\noindent Next, we choose an $8$-length control skipping pattern, $\rho_1 = 11010011$ that satisfies  $r_{min}=0.5$. With the same choice of attack vector as used earlier in the periodic execution, this time, running the system with the pattern $\rho_1$, we observe the following cases.

\par \textbf{O1}. While the $11$-length attack could drive the system to an unsafe state and remain stealthy for fully periodic execution, in case of execution with the pattern $\rho_1$, it is detected at \emph{9-th} iteration just after driving the system to an unsafe state at \emph{8-th} iteration. This happens because due to the control skips the attacker's efforts in those samples are not affecting the system. 
This leads to better unbiased estimation in such iterations which may  create a large residue resulting detection in future iterations that are under attack.
\par \textbf{O2}. We also find that no successful but stealthy attack of length $d < 15$ is possible for this system running with the  pattern $\rho_1$. System response for this pattern-based execution ($\rho_1^{\omega}$) of the system, with a successful and stealthy attack vector of length $d=15$ is depicted in Fig.~\ref{fig:attackOnPattaern1}. 
 The control skips reduce the amount of attack that could have been injected while remaining stealthy.
 In general, the value of $d_{min}$ is dependent on the choice of pattern because system-behaviour under a control skipping pattern depends on the positions of the control skips and the nature of the system. 
    
\par \textbf{O3}. Since the minimum attack-length $d_{min}=15$ in this case, we can set $n_{down}=14$  increasing the attack resilience (i.e., $n_{down}/n_{up}$) by $\approx 30$\% in comparison with the periodic execution ($n_{down}=10$) for a fixed value of $n_{up}$. This increment in $n_{down}$ in-turns reduces the computation time of the IDS saving the resource bandwidth.

\par \textbf{O4}. In general, there may exist multiple patterns that are equally resilient (i.e. with similar $d_{min}$). 
Among such patterns with similar resilience, it makes sense to choose the one providing better control performance, e.g. lower LQR cost in our setting. 
The observations indicate that it is possible for a CPS to  be more resilient to false data injection attacks when running with a control skipping pattern when compared with fully periodic execution. However, we need an efficient algorithmic framework in order to search for such performance preserving attack resilient patterns. The next section describes such a framework in detail. 



\section{Proposed Methodology}
\label{sec:design_analysis}
As motivated earlier, our framework has two distinct steps which are discussed next. 
\par  $\bullet$ \textit{Step-1:} We synthesize a set of control skipping patterns $\mathcal{P}$, and rank them according to their control performance (ref. Algo.~\ref{alg:Patternprune}). This helps in filtering upfront all the patterns that violate the desired control performance.
\par  $\bullet$ \textit{Step-2:} In this step, we synthesize the set $\mathcal{P}_R \subset \mathcal{P}$ of  most attack resilient control skipping pattern(s), which guarantee minimal resource usage and best ranked control performance (ref. Algo.~\ref{alg:ids}). For this, we  compute successful yet stealthy attack vectors for control schedules $\in \mathcal{P}$ using  Algo.~\ref{alg:PatternBasedExecution}.  


\subsection{Synthesizing and Ranking Patterns based on Control Performance}\label{subsec:pattern_prune}
Recall that in Sec.~\ref{subsubsec:stability}, we  already limit the number of allowable execution skips  by imposing the constraint of \textit{minimum execution rate} $r_{min}$. Yet, for a large $l$, the  number of patterns is still very large for testing attack resilience. 
Moreover w.r.t. resilience, it is important to remember that a pattern under repeated execution is equivalent to any of its cyclic shifts as the attack can start at any point of execution. For example, the patterns 1110 and 0111 are equivalent since one can be derived from another using cyclic shifts. Both represent the same infinite control schedule, i.e. $(1110)^\omega=(0111)^\omega$. Thus, we need a measure which i) considers any two patterns that are cyclic shifts of each other as equivalent and  ii) is also indicative of the control performance of a candidate pattern. Using such a measure to rank patterns provides the following advantage. Patterns whose cyclic shifts  are already tested for resilience need not be tested again thus eliminating expensive computation. Since the computation of control performance index $J$ for every pattern requires evaluating a complex quadratic expression, a lightweight equivalent index can help in ranking of patterns w.r.t. performance thus ensuring that our method outputs patterns which are both resilient as well as performance preserving.
\par In order to model these strategies in the pattern-synthesis approach, we use the correlation between the structure of a pattern (i.e. relative position of skips) and its LQR cost. A well known theory ~\cite{jia2007graceful} establishes that  \textit{a pattern having most uniform execution skips (i.e., uniform distribution of \lq $0$\rq s) exhibits the lowest LQR cost}. The uniformity of binary patterns is defined in literature using the notion of  {\em upper mechanical binary  word} \cite{choffrut1997combinatorics}. 
Following the same we can define the notion of  \textit{most uniform skipping pattern} as follows. 
\begin{Definition}
\label{def:uniform_pattern}
  \textbf{Uniform Control Skipping Pattern:} An $l$-length control skipping pattern $\rho$ with number of control execution skips = $\floor{(1-r_{min})\times l}$ (where $r_{min}$ is as discussed earlier), is considered to be {\em fully uniform} when the number of skips in each overlapping $q$-length sub-sequence of $\rho$ is exactly one, where $q=\lceil l/\{l\times (1-r_{min})\}\rceil$.\hfill $\Box$ 
\end{Definition}
For example, consider $\rho_1=101010$ and $\rho_2=111000$ which are $l=6$ length patterns  satisfying $r_{min}=0.5$. We can claim $\rho_1$ is a uniform pattern but $\rho_2$ is not. Because, all $6$ overlapping $q= 6/3= 2$ length sub-sequences  $\tupleangle{10,01,10,01,10,01}$ of $\rho_1$ contain  exactly one \lq $0$\rq (i.e., skip), whereas $\rho_2$ has only $2$ such sub-sequences ($3$-rd and $6$-th) among  $\tupleangle{11,11,\underline{10},00,00,\underline{01}}$. It is understood that the trailing  sub-sequences are derived by cyclic rotation of the pattern. With this observation we try to rank the patterns of a given length based on how much they deviate from absolute uniformity. 
For this we define a function $skipCount(\rho, i,q)$ which outputs the number of \lq $0$\rq s in a  $q$-length sub-sequence of $\rho$ starting from $\rho[i]$ (i.e. the $i$-th term of $\rho$). In case $i+q-1 > l$, the sub-sequence under consideration will wrap back to the front, i.e., it will be given by $\rho[i]\rho[i+1]\cdots\rho[l-1]\cdots\rho[(i+q-1)\%l]$. 
Based on this, we employ the following metric of {\em non-uniformity} for patterns in our work. 
\begin{Definition}
\label{def:lrq_dist}
\textbf{\texttt{LQR-Distance()}: } Consider an $l$-length control skipping pattern $\rho$ with   minimum execution rate $r_{min}$ and $q=\lceil l/\{l\times (1-r_{min})\}\rceil$. For a given $\rho$, the index \texttt{LQR-Distance}($\rho$) is defined as, 
\texttt{ LQR-Distance}$(\rho)=\sum_{i=0}^{l-1}|min(0,skipCount(\rho,i,q)-1)|.$
Given patterns $\rho_i$ and $\rho_j$, $\rho_i$ is considered more non-uniform than $\rho_j$ if \texttt{LQR-Distance($\rho_i$)} > \texttt{LQR-Distance($\rho_j$)}.\hfill $\Box$
\end{Definition}
\noindent The subtraction of 1 is done since $skipCount$ is expected as `1' in all cases for  perfect uniformity.
Considering the previously used patterns $\rho_1=101010,\rho_2=111000$ ($l=6,q=2$), we have  
$\texttt{LQR-Distance}(\rho_1)=0$, and 
$\texttt{LQR-Distance}(\rho_2)=2$. This gives a measure that among these two patterns with same amount of skip, the control performance of $\rho_1$ will be better. Also, the measure will be same for all cyclic shifts of a pattern since the definition itself accounts for it. In that way, all patterns 
with equal length and equal number of execution skips
which evaluate to same value of $\texttt{LQR-Distance}$ will be similar in control performance.

\par We use the measure defined to rank and classify patterns in Algorithm  \ref{alg:Patternprune}. In this algorithm, we consider a user specified pattern length $l$ and number of skips fixed as $\theta$. With this,  we first generate all possible $l$-length patterns  (Line  \ref{alg:pattaernprune:allpat}) with $\theta$ skips. Then we group patterns with same  \texttt{LQR-Distance} in the same set (Line \ref{alg:pattaernprune:grp}). Patterns with cyclic equivalence get automatically grouped with same \texttt{LQR-Distance} value in the data structures called pattern-lists denoted by $s \in  \mathcal{S}_l$  where $\mathcal{S}_l$ is the overall collection of $l$ length patterns with $\theta$ skips. For  patterns with same  \texttt{LQR-Distance}, i.e., in the same pattern-list, we carry out the following pruning operation. For any  $\rho \in s$, we eliminate all other patterns which are cyclic shift equivalent of $\rho$ (Line \ref{alg:pattaernprune:omitcycles}) since  they are equivalent w.r.t. both performance  as well as resilience (as we shall see). 
Next, we include this pruned set in  $\mathcal{S}^{\prime}$ (Line \ref{alg:pattaernprune:addset}). After the pruning is completed for each pattern-list, the  collective set is inserted into the final set of patterns $\mathcal{P}$ (Line \ref{alg:pattaernprune:addtop}). 

\begin{algorithm}\scriptsize
\caption{Performance-based Pattern Synthesis}
\label{alg:Patternprune}
\begin{algorithmic}[1]
\Require 
pattern length: $l$,Exact no. of skips: $\theta$,
\Ensure{Set of pattern, $\mathcal{P}$, sorted in ascending order of the LQR cost}
\Function{Rank\_Pattern}{$l,\theta$}
\State \label{alg:pattaernprune:setinit}$\mathcal{P},\mathcal{S}^{\prime},\bar{\mathcal{S}}_l\gets \Phi$;\Comment{Initialized with NULL}
\State \label{alg:pattaernprune:allpat}$\mathcal{S}_{l} \gets$ all possible $l$ length pattern with $l-\theta$ number of \q{$1$}s;\Comment{performance criteria}
\State \label{alg:pattaernprune:q}$q\gets \ceil{ \{l/\theta\}}$;
\Comment{the uniformity condition}
\For{each pattern $\rho\in\mathcal{S}_{l}$}
\State \label{alg:pattaernprune:grp}$\bar{\mathcal{S}}_l[$\texttt{LQR-Distance}$(\rho,q)]\gets \rho$ \Comment{group patterns w.r.t. $\texttt{LQR-Distance}$}
\EndFor
\For{each pattern-list $\mathbf{s} \in\bar{\mathcal{S}}_l$}

\For{each $\rho \in \mathbf{s}$}
\For{each cyclic shift $\rho'\in \mathbf{s}$}\label{alg:pattaernprune:allcycles}
\State \label{alg:pattaernprune:omitcycles}$\mathbf{s}\gets \mathbf{s} \setminus \rho'$\Comment{omitting cyclic shifts of $\rho$}
\EndFor
\EndFor
\State \label{alg:pattaernprune:addset}$\mathcal{S}^{\prime}\gets\mathcal{S}^{\prime}\cup \mathbf{s}$
\EndFor

\State \label{alg:pattaernprune:addtop}$\mathcal{P}\gets\mathcal{P}\cup \mathcal{S}^\prime $;
\State \label{alg:pattaernprune:returnp}\Return $\mathcal{P}$; 
\EndFunction
%
\end{algorithmic}
\end{algorithm}
%
\subsection{Attack Vector Synthesis}\label{subsec:attack_synth}
In order to synthesize patterns having best  attack-resilience, an important step is to verify the existence of {\em successful} and {\em stealthy} attack vectors for patterns under test. We develop a formal approach to synthesize attack vectors for control skipping patterns as outlined in Algorithm  \ref{alg:PatternBasedExecution}. We build on earlier work on attack vector synthesis for periodic controllers \cite{ipsita2020}. 

\begin{algorithm}\scriptsize
	\caption{Attack Vector Synthesis for Pattern-based Execution}
	\label{alg:PatternBasedExecution}
	\begin{algorithmic}[1]
		\Require{Attack length: $d$, pattern: $\rho$, IDS up-time: $n_{up}$, detector threshold: $Th$, inner safety region: $\mathcal{C}_1$, outer safety region: $\mathcal{C}_2$}
		\Ensure{Attack vector $\mathcal{A}_{d}$ of length $d$ (if it exists, otherwise NULL)}
		\Function{SynAttVec}{$d,\rho,n_{up},Th$}
		\State \label{alg:stateinit}$x[0] \in \mathcal{C}_1$; $\hat{x}[0]\gets0$; $u[0]\gets K\hat{x}[0]\gets0$; $y[0]\gets Cx[0]$;\Comment{Starting from $\mathcal{C}_1$}
		\State \label{alg:resinit}$r[0]\gets y[0]-C\hat{x}[0];\tilde{u}[0]\gets u[0];\tilde{y}[0]\gets y[0];$
		\For{$k=1$ to $d+n_{up}$}
		\State \label{alg:stateprog} $x[k] \gets Ax[k-1] + B\tilde{u}[k-1]$; $\hat{x}[k] \gets A\hat{x}[k-1] + Bu[k-1] + Lr[k-1]$;
		\If{$k\leq d$}\label{alg:ifinatklen}
		$\triangle u[k] \gets$ \textbf{nondet()};\;$\triangle y[k] \gets$ \textbf{nondet()};
		\Else{}\label{alg:elsenoatk}
		$\triangle u[k] \gets 0$; $\triangle y[k] \gets 0$;
		\EndIf
		\If{$\rho[k] = 1$}\label{alg:nodropu}
		$u[k]\gets K\hat{x}[k];$
		$\tilde{u}[k]\gets u[k]+\triangle u[k];$
		\Else{}\label{alg:dropu}
		$u[k]\gets u[k-1];\;\tilde{u}[k]\gets\tilde{u}[k-1]$;\Comment{Skip Execution}
		\EndIf
		\State \label{alg:residueprog}$\tilde{y}[k]\gets y[k]+\triangle y[k]$; $r[k]\gets\tilde{y}[k]-C\hat{x}[k];$
		\EndFor
        \State \label{alg:assert}$\Phi \gets$\textbf{assert}(($|r[1]|\leq Th \wedge.. |r[d+n_{up}]|\leq Th$) $\wedge $ ($x[1]\notin \mathcal{C}_2 \vee .. \vee x[d+n_{up}]\notin \mathcal{C}_2$));
		\If {$\Phi$ is {\em $unsatisfiable$}} 
		\Return NULL;
		\Else $\;$\label{alg:retatkvec}
		\Return $\mathcal{A}_d \gets \begin{bmatrix}
		\triangle u_1 & \cdots & \triangle u_d\\ 
		\triangle y_1 & \cdots & \triangle y_d
		\end{bmatrix};$
		\EndIf
		\EndFunction
	\end{algorithmic}
\end{algorithm}
\noindent The function \textsc{SynAttVec} in Algo.~\ref{alg:PatternBasedExecution}, {\em symbolically} executes the system starting from any initial state $x[0]$ inside the inner safety region $\mathcal{C}_1$(Line \ref{alg:stateinit}) for  $d+n_{up}$ control samples  following Eqn. \eqref{eq:state_space}. In each sample $k$, we introduce two non-deterministic variables $\triangle u[k]$ and $\triangle y[k]$ to model the actuation and measurement errors introduced by the adversary (Line \ref{alg:ifinatklen}). Attack length is bounded to $d$ by setting these variables to zero for each iteration $k > d$. In case of the skip in $k$-th control execution (i.e., $\rho[k]=0$), $x[k],r[k],y[k]$ are calculated following Eq. \eqref{eq:loop_matrix_drop} ($u[k], \tilde{u}[k]$ are updated using the last calculated $u[k-1],\tilde{u}[k-1]$ , in line \ref{alg:dropu}). The function at the end validates an assertion using the SMT solver Z3 \cite{de2008z3} to check if any attack of length $d$ that is stealthy over $d+n_{up}$ samples (i.e., until further activation of IDS), violates the safety requirements of the system in any control sample (Line \ref{alg:assert}). On getting satisfiable solution from the solver, \textsc{SynAttVec}() returns a successful attack vector $\mathcal{A}_d$ of length $d$ (Line \ref{alg:retatkvec}). Otherwise it returns NULL. This guarantees that no attack vector of length $d$ exists that remains stealthy over $d+n_{up}$ control samples and successfully violates the safety of the system in any of those samples. 

\subsection{Synthesizing Attack Resilient Patterns} 
\label{subsec:sporadic_ids}
As described earlier, given a control system, we compute a reduced set $\mathcal{P}$ with fixed length control skipping patterns and fixed number of skips, ranked according to their control performance using Algo.~\ref{alg:Patternprune}. We  use the set $\mathcal{P}$ to find a further  pruned set of patterns $\mathcal{P}_l \subset \mathcal{P}$ where each $\rho \in \mathcal{P}_l$ has a sporadic IDS specification $\tupleangle{n_{up}^\rho$, $n_{down}^\rho}$ for a detector threshold $Th$, ensuring the following.\\
\textbf{1}) The ranking of the $l$ length patterns (w.r.t. descending order of Quality of Control (QoC)) as set by Algo.~\ref{alg:Patternprune} is  maintained in $\mathcal{P}_l$.\\
\textbf{2}) Given the inner and outer safety regions, $\mathcal{C}_1$ and $\mathcal{C}_2$, (ref. Fig. \ref{fig:ids}), starting from anywhere inside $\mathcal{C}_2$, the system will reach  $\mathcal{C}_1$ under a safe scenario with no stealthy attack as guaranteed by an IDS within $n_{up}^\rho$ iterations.\\
\textbf{3}) $(n_{down}^\rho+1)$ is minimum attack length required to drive the system to an unsafe state while remaining stealthy.\\
\textbf{4}) Attack resilience, i.e., $(n_{down}^\rho/n_{up}^\rho)$ will be maximum and same for all the patterns in $\mathcal{P}_l$ ensuring minimum IDS execution rate, i.e., $n_{up}^\rho/(n_{down}^\rho+n_{up}^\rho)$.\\ 
We derive such a set $\mathcal{P}_l$ for all allowable number of skips $\theta \in [1,\lfloor l\times (1-r_{min})\rfloor ]$ (for certain $l$ length) and arrange them in increasing order of control skips. 
\noindent The method is outlined in Algo.~\ref{alg:ids}. 
Here, our goal is to output set of patterns, $\mathcal{P}_R$, with most attack resilience that would help us design better sporadic IDS schemes with provable security, improved resource utilization ensuring best performance. We define $\rho^{*}=1$ as the $1$-length pattern representing the periodic execution, i.e., $(\rho^{*})^\omega=1^\omega$ in order to represent existing IDS schemes in literature.  
\begin{algorithm}[t]\scriptsize
	\caption{Most Attack Resilient Pattern Synthesis}
	\label{alg:ids}
	\begin{algorithmic}[1]
		\Require{Desired pattern length $l$,
		$1$-length pattern for periodic control execution: $\rho^{*}$, 
		detector Threshold: $Th$, inner and outer safety regions: $\mathcal{C}_1$ and $\mathcal{C}_2$}, plant and controller matrices: $A, B, C, K$, Min. execution rate: $r_{min}$
		\Ensure{Set of most attack resilient $l$ length patterns $\mathcal{P}_R$}
		\State \label{alg:ids:skipmax}$\theta_{max}\gets \floor{l\times(1-r_{min})}$;\Comment{initializing with maximum number of skips allowed}
		\For{each $\theta \in [1,\theta_{max}]$}\label{alg:ids:foreachskip}
		\State \label{alg:ids:deriveP}$\mathcal{P}\gets \Call{RankPattern}{l,\theta}$;
		\State \label{alg:ids:periodicuptime} $n_{up}^{\rho^{*}} \gets$\Call{FindOnTime}{$\rho^{*},\mathcal{C}_1,\mathcal{C}_2$}; \ $d \gets 1$; \Comment{Finding ontime of the IDS, Initializing $d$}
		\State \label{alg:ids:dmin} $d_{min}^{*} \gets$\Call{MinAttLen}{$\rho^{*}, d, n_{up}^{\rho^{*}}, Th$};
		\State \label{alg:ids:minndown}$n_{down}^{\rho}\gets d^{*}_{min}-1$; $d_{min}\gets d_{min}^{*}$\Comment{Initializing with minimum attack length for $1^{\omega}$}
		\State \label{alg:ids:minidsrate}$rate_{\rho^{*}} \gets n_{up}^{\rho^{*}}/(n_{down}^{\rho^{*}}+n_{up}^{\rho^{*}})$; \ $rate_{min} \gets rate_{\rho^{*}}$;\Comment{Initializing with $1^{\omega}$ IDS rate}
		\For{each pattern $\rho\in\mathcal{P}$}
		\State \label{alg:ids:updowntime} $n_{up}^\rho \gets$\Call{FindOnTime}{$\rho,\mathcal{C}_1,\mathcal{C}_2$};\ $d \gets$ \Call{MinAttLen}{$\rho, d_{min}, n_{up}^\rho, Th$};
		\If{ $d \geq d_{min}$} \label{alg:ids:ifbetterpattern}
		\State \label{alg:ids:updaterate}$n_{down}^\rho \gets$ $d-1$; \ $rate_\rho \gets n_{up}^\rho/(n_{down}^\rho+n_{up}^\rho)$;
		\If{$rate_\rho > rate_{min}$}\label{alg:ids:ifbetterrate} $\mathcal{P} \gets \mathcal{P}\setminus\rho$
		\Else $\;rate_{min} \gets rate_\rho;\ d_{min}\gets d$;\label{alg:ids:addgoodpattern}
		\EndIf
		\Else $\;\mathcal{P} \gets \mathcal{P} \setminus \rho $\label{alg:ids:omitbadpattern}
		\EndIf
		\EndFor
		\For{each pattern $\rho \in \mathcal{P}$}
		\If{$rate_{\rho} == rate_{min}$}\label{alg:ids:minratetop} $\mathcal{P}_l \gets \rho$;
		\EndIf
		\EndFor
		\State \label{alg:ids:sortpbyskip}$\mathcal{P}_R[\theta]\gets \mathcal{P}_l$\Comment{Store $l$ length most attack resilient patterns performance wise}
		\EndFor
		\State \Return \label{alg:ids:returnp}$\mathcal{P}_R$
		\Function{MinAttLen}{$\rho$, $d_{m}$, $n_{up}$, $Th$}\label{alg:ids:minatklenfunc}
		\State \label{alg:ids:initdmin}$d\gets d_{m};$
		\Repeat \label{alg:ids:incrd}
		$d\gets d+1$
		\For {$i=0$ to $|\rho|-1$}
		\State \label{alg:ids:checkallcycle}$\rho^\prime \gets$ $i$-times left cyclic shift of pattern $\rho$;
		\If{\Call{SynAttVec}{$d,\rho^\prime,n_{up},Th$} $\neq NULL$} \Return $d-1$;\label{alg:ids:ifatkvecfound}
		\EndIf
		\EndFor
		\Until{\Call{SynAttVec}{$d,\rho^\prime,n_{up},Th$}$ = NULL$}\label{alg:ids:rptuntillnoatkvec}
		\EndFunction
		\Function{FindOnTime}{$\rho,\mathcal{C}_1,\mathcal{C}_2$}\label{alg:ids:findontimefunc}
		\State \label{inituptime}$n\gets 1$
		\For {$i=0$ to $|\rho|-1$}
		\State \label{alg:ids:onallcycles}$\rho^\prime \gets$ $i$-times left cyclic shift of pattern $\rho$;
		\Repeat\label{alg:ids:rptnup}
		\State \label{alg:ids:sysinit} $x[0] \in \mathcal{C}_2;\;u[0]=0;\;r[0]=0;$
		\For{$k=1$ to $n$}
		\State \label{alg:ids:residue} $r[k-1]\gets{y}[k-1]-C\hat{x}[k-1]$;
		\State \label{alg:ids:states}$\hat{x}[k] \gets A\hat{x}[k-1] + Bu[k-1] + Lr[k-1]$; $x[k] \gets Ax[k-1] + Bu[k-1]$;
		\If{$\rho'[k] = 1$} $u[k]= K\hat{x}[k]$;\label{alg:ids:nodropu}
		\Else{} \label{alg:ids:dropu}$u[k]\gets u[k-1]$;\Comment{Skip Execution}
		\EndIf
		\EndFor
		\State \label{alg:ids:assert} $\Phi\gets$ {\bf assert($|r[1]|\leq Th \wedge\cdots\wedge |r[n]|\leq Th \wedge x[n]\not\in \mathcal{C}_1$)}; 
		\State \label{alg:ids:incrnup} $n\gets n+1$
		\Until{$\Phi$ is $unsatisfiable$}\label{alg:ids:ifuunsat}
		\State \label{alg:ids:decrnup}$n\gets n-1$
		\EndFor
		\State \label{alg:ids:retnup}\Return $n$
		\EndFunction
	\end{algorithmic}
\end{algorithm}

\par In Algo.~\ref{alg:ids}, we compute IDS up and down time for any pattern using \textsc{FindOnTime}() and  \textsc{MinAttLen}() function respectively.
\textsc{FindOnTime}() returns the minimum number of iterations required by following the pattern $\rho$ to formally guarantee that the system starting from any state $x[k]$ in the outer safety region $\mathcal{C}_2$ (as a result of successful attack) will be in a state inside the inner safety region $\mathcal{C}_1$ (Lines \ref{alg:ids:findontimefunc}-\ref{alg:ids:retnup}). We symbolically simulate attack-free closed loop iterations of the system starting from an initial state $x[0] \in \mathcal{C}_2$ according to the pattern $\rho^\prime$ (where $\rho^\prime$ represents a left cyclic shift of the pattern $\rho$) (Lines \ref{alg:ids:onallcycles}-\ref{alg:ids:sysinit}). We use the clause $x[k]\not\in \mathcal{C}_1$ which implies that the system is not inside the inner  safety region {$\mathcal{C}_1$} after $k$ iterations (Line \ref{alg:ids:assert}). This assertion is the negation of our design requirement for the up-time $n_{up}$ of the IDS. 
If the assertion $\Phi$ is found to be unsatisfiable using SMT solver, then our design requirement is valid (Line \ref{alg:ids:ifuunsat}-\ref{alg:ids:decrnup}). However, if $\Phi$ is satisfiable, then we infer that the present IDS up-time, $n$, is not sufficient to bring the system to the inner safety region $\mathcal{C}_1$ starting from any point in the outer safety region $\mathcal{C}_2$, and we increase $n$ until $\Phi$ becomes unsatisfiable (Line \ref{alg:ids:incrnup}). We now repeat this procedure to find the maximum 
value of $n$
that satisfies our design requirement over all possible cyclic shifts of the pattern $\rho$ (Lines \ref{alg:ids:rptnup}-\ref{alg:ids:incrnup}). We check for all possible such shifts since the system can start from $\mathcal{C}_2$ while executing any position in the pattern. The value of $n$ thus  found is a safe up-time of the sporadic IDS designed using an attack resilient control skipping pattern $\rho$, i.e. $n_{up}^{\rho}=n$ (Line \ref{alg:ids:retnup}).

\par The \textsc{MinAttLen}() function on the other hand computes all possible cyclic shifts of the input pattern as $\rho^\prime$ (Line \ref{alg:ids:checkallcycle}) and calls the function \textsc{SynAttVec}() (Line \ref{alg:ids:ifatkvecfound}) which checks for existence of possible stealthy and successful attack vector of length $d$ (initialized with input length $d_{m}$ in line \ref{alg:ids:initdmin}). 
If no attack vector of length $d$ exists, we can claim that the system can not be made unsafe with stealthy attack of length $d$. Hence, we search again for an attack vector by increasing the attack length by $1$ (Line \ref{alg:ids:incrd}). Otherwise, on finding a successful and stealthy attack vector of $d$ length, we terminate by decreasing the length by $1$ and return the length as minimum attack length (Line \ref{alg:ids:ifatkvecfound}). 

\par We start Algo.~\ref{alg:ids} by choosing a certain number of control skips $\theta \leq \theta_{max}$,which is the maximum number of allowed control skips for $l$ length pattern, calculated using the length input $l$ and minimum execution rate criteria for a system i.e. $r_{min}$ ($\theta_{max}=\lfloor l\times (1- r_{min})\rfloor$, Lines \ref{alg:ids:skipmax}- \ref{alg:ids:foreachskip}). For this $\tupleangle{l,\theta}$ pair we call $\Call{RankPattern}{l,\theta}$ to get the pruned and  Quality of Control (QoC) wise ordered set of $l$ length patterns $\mathcal{P}$. Our aim here is to make the IDS scheme as much  sporadic as possible i.e.  minimizing the IDS execution rate ($n_{up}/(n_{up}+n_{down})$)) w.r.t. their periodic counterpart by examining all $l$ length patterns. So we start our attack resilience analysis with the periodic pattern $\rho^{*}$. We derive $d_{min}^{*}$ i.e., minimum attack length for $\rho^{*}$ (periodic execution) and update $d_{min}$ with it first. Then we calculate down time for $\rho^{*}$, i.e., $n_{down}^{\rho^{*}}=d_{min}^{*}-1$ (Line \ref{alg:ids:dmin}-\ref{alg:ids:minndown}). We compute IDS up-time for $\rho^{*}$ in line \ref{alg:ids:periodicuptime}. Then we initialize $rate_{min}$ with IDS execution rate for periodic execution i.e., $rate_{\rho^{*}}$ (Line \ref{alg:ids:minidsrate}). Next, for every pattern $\rho \in \mathcal{P}$, we first calculate the up-time ($n_{up}^{\rho}$) and minimum attack length ($d$) for $\rho$ using the functions \textsc{FindOnTime}()  and \textsc{MinAttLen}() respectively (Line \ref{alg:ids:updowntime}). If $d$ is larger than or equal to $d_{min}$ indicating better attack resilience ($n^{\rho}_{down}/n^{\rho}_{up}$) than last found most attack resilient pattern (Line \ref{alg:ids:ifbetterpattern}), we compute $rate_\rho$, the execution rate for the pattern $\rho$ (Line \ref{alg:ids:updaterate}). 
A pattern $\rho$ is removed from $\mathcal{P}$ if $rate_\rho>rate_{min}$ since $\rho$ can not reduce IDS utilization when compared to last found best candidate (Line \ref{alg:ids:ifbetterrate}). Otherwise, $rate_{min}$ and $d_{min}$ are updated with  $rate_{\rho}$ and $d$ respectively (Line \ref{alg:ids:addgoodpattern}).
\par While repeating the above procedure for all patterns ($\forall \rho \in \mathcal{P}$), we pick the patterns with least IDS execution rate $rate_{min}$ from $\mathcal{P}$ and insert them into $\mathcal{P}_l$ maintaining their actual order (Line \ref{alg:ids:minratetop}). This sorted set $\mathcal{P}_l$ has following properties, i.e. $\forall \rho \in \mathcal{P}_l$, \textbf{(i)} $rate_{\rho} = rate_{min}$ among all $l$ length patterns with $\theta$ number of skips  and \textbf{(ii)} all patterns in $\mathcal{P}_l$ are sorted in increasing order of \texttt{LQR-Distance}. We store $\mathcal{P}_l$ derived for all possible skips ($\leq \theta_{max}$) for a fixed length $l$ in $\mathcal{P}_R$, indexing them with number of skips (Line \ref{alg:ids:sortpbyskip}) and finally returning this set (Line \ref{alg:ids:returnp}). In $\mathcal{P}_R$, the set of patterns with smaller number of skips are better in control performance and patterns with same number of skips are internally sorted (in each entry of $\mathcal{P}_R$) following uniformity measure. 
A system running with any of the $l$ length control skipping patterns $\in \mathcal{P}_R$ meets the desired performance criteria with best QoC and a sporadic IDS can be designed for this system  
having a formal guarantee of the security against false data injection attack with minimum IDS activation. 



\section{Results} 
\label{sec:results}
We demonstrate the efficacy of our proposed approach considering two systems from the automotive domain. The systems are \textit{Vehicle Dynamic Controller} (VDC) and \textit{Trajectory Tracking Controller} (TTC).
\subsection{Case Studies}
\label{subsec:case studies}
VDC regulates the lateral dynamics of a vehicle  by controlling its side slip ($\beta$) and yaw rate ($\gamma$)~\cite{zheng2006controller}. The  control input in this case is the steering angle. For TTC  \cite{jovanov2017sporadic}, details about the system specifications are given in Sec.~\ref{sec:motivating_example}. For both the systems, system matrices ($A,B,C$), sampling period ($h$), outer ($\mathcal{C}_2$), inner ($\mathcal{C}_1$) safety regions of the state variables and detector thresholds ($Th$) are  given in Tab.~\ref{tab:sysSpecs}. Safety regions are determined following \cite{yawrateDatasheet,steeringAngleDatasheet}. 
\begin{table}[!h]
\centering
\vspace{1pt}
\scriptsize
\caption{System Specifications}
\label{tab:sysSpecs}
\vspace{-1.5em}
\begin{tabular}{|l|l|l|l|l|}
\hline
\rowcolor[HTML]{C0C0C0} 
\multicolumn{1}{|c|}{\cellcolor[HTML]{C0C0C0}System} & \multicolumn{1}{c|}{\cellcolor[HTML]{C0C0C0}Specifications} & \multicolumn{1}{c|}{\cellcolor[HTML]{C0C0C0}$\mathcal{C}_2$} & \multicolumn{1}{c|}{\cellcolor[HTML]{C0C0C0}$\mathcal{C}_1$} &
\multicolumn{1}{c|}{\cellcolor[HTML]{C0C0C0}$Th$}\\
\hline
VDC & \begin{tabular}[c]{@{}l@{}}A = {[}0.4450,-0.0458;1.2939,0.4402{]};\\
B = {[}0.0550;4.5607{]}; C = {[}0,1{]};\\
h = 0.1sec; K = {[}-0.0987;0.1420{]};\\
L = {[}-0.0390;0.4339{]}
\end{tabular} &
\begin{tabular}
[c]{@{}l@{}}$\beta \in$ {[}-1, 1{]}\\
$\gamma \in$ {[}-2, 2{]}
\end{tabular} &
\begin{tabular}
[c]{@{}l@{}}$\beta \in$ {[}-0.1, 0.1{]}\\ 
$\gamma \in$ {[}-0.2, 0.2{]}
\end{tabular} &
0.003\\ 
\hline
\begin{tabular}
[c]{@{}l@{}}TTC\end{tabular} &
\begin{tabular}
[c]{@{}l@{}}A = {[}1.0000, 0.1000;0, 1.0000{]};\\ 
B = {[}0.0050;0.1000{]}; C = {[}1 0{]};\\ 
h = 0.1sec; K = {[}16.0302,    5.6622{]};\\
L = {[}1.8721;9.6532{]}\end{tabular} & \begin{tabular}[c]{@{}l@{}}$D \in$ {[}-25, 25{]}\\
$V \in$ {[}-30, 30{]}\end{tabular} &
\begin{tabular}
[c]{@{}l@{}}$D \in$ {[}-15, 15{]}\\
$V \in$ {[}-18, 18{]}\end{tabular} &
2\\
\hline
\end{tabular}
\end{table}

\begin{table}
\centering
\footnotesize
\vspace{1.5pt}
\caption{Designed Sporadic IDS schemes for VDC and TTC}
\vspace{-0.8em}
\begin{tabular}{|c|c|c|c|c|c|}
\hline
\rowcolor[HTML]{C0C0C0} 
\textbf{Sys.}                                           & $\tupleangle{\mathbf{l,\theta}}$        & \textbf{pattern}     & $\tupleangle{\mathbf{n_{down},n_{up}}}$ & $\mathbf{rate}$ & \textbf{\texttt{LQR-D}} \\ \hline
\cellcolor[HTML]{C0C0C0}                                &        -               & 1          & 10,3               & 0.2308          & -             \\ \cline{2-6} 
\cellcolor[HTML]{C0C0C0}  & {10,3} & \textbf{1010011111}  & \textbf{15,3}      & \textbf{0.1667} & \textbf{3}    \\ \cline{2-6} 
\cellcolor[HTML]{C0C0C0}                                & 10,4                   & 1101011100           & 14,3               & 0.1765          & 1             \\ \cline{2-6} 
\cellcolor[HTML]{C0C0C0}                                & 10,5                   & 1101001010           & 13,3               & 0.1875          & 1             \\\cline{2-6} 
\cellcolor[HTML]{C0C0C0}                                & {11,4} & \textbf{11010111100} & \textbf{15,3}      & \textbf{0.1667} & \textbf{2}    \\ \cline{2-6}
\cellcolor[HTML]{C0C0C0}                                &                        & 10100101011          & 13,3               & 0.1875          & 1             \\ \cline{3-6} 
\multirow{-6}{*}{\cellcolor[HTML]{C0C0C0}\textbf{TTC}}                                & \multirow{-2}{*}{11,5} & 10100111010          & 13,3               & 0.1875          & 1             \\  
\hline\hline 

\cellcolor[HTML]{C0C0C0}                                &        -               & 1          & 2,3               & 0.6          & -             \\ \cline{2-6} 
\multirow{10}{*}{\cellcolor[HTML]{C0C0C0}\textbf{VDC}} & 2,1                    & \textbf{10}          & \textbf{5,3}       & \textbf{0.375} & \textbf{0}    \\ \cline{2-6} 
\cellcolor[HTML]{C0C0C0}                                & 5,2                    & 11010                & 4,3                & 0.4286          & 0             \\ \cline{2-6} 
\cellcolor[HTML]{C0C0C0}                              &                        & \textbf{110010}      & \textbf{5,3}       & \textbf{0.375} & \textbf{1}    \\ \cline{3-6} 
\cellcolor[HTML]{C0C0C0}                                &   & \textbf{110100}      & \textbf{5,3}       & \textbf{0.375} & \textbf{1}    \\ \cline{3-6} 
\cellcolor[HTML]{C0C0C0}                                 &             \multirow{-3}{*}{6,3}            & \textbf{100011}      & \textbf{5,3}       & \textbf{0.375} & \textbf{2}    \\ \cline{2-6} 
\cellcolor[HTML]{C0C0C0}                                &\multirow{3}{*}{10,5}& 1100101010           & 4,3                & 0.4286          & 1             \\ \cline{3-6} 
\cellcolor[HTML]{C0C0C0}                                &  & 1101001010           & 4,3                & 0.4286          & 1             \\ \cline{3-6} 
\cellcolor[HTML]{C0C0C0}                                &                        & 1000111001           & 4,3                & 0.4286          & 2             \\ \cline{2-6}
\cellcolor[HTML]{C0C0C0}                                &                        & 110001110010         & 4,3                & 0.4286          & 3             \\ \cline{3-6} 
\multirow{-10}{*}{\cellcolor[HTML]{C0C0C0}\textbf{VDC}}                                & \multirow{-2}{*}{12,6} & 110100111000         & 4,3                & 0.4286          & 3             \\ \hline
\end{tabular}
\label{tab:IDSrate}

\end{table}

%
For the above systems, we first report in Row 1 of both parts of  Tab.~\ref{tab:IDSrate}  the results for sporadic IDS design with fully periodic execution ($1^\omega$) similar to  \cite{jovanov2017sporadic}. For periodic execution, our method computes IDS up-time $n_{up}=3, 3$, and minimum attack length $d_{min} = 11, 3$ for TTC and VDC respectively. These are given in Row 1, Col. 4  of both parts in  Tab.~\ref{tab:IDSrate} ($n_{down}=d_{min}-1$). Using these, IDS execution rates ($rate$) of periodic execution are calculated and reported in Col. 5 of Tab.~\ref{tab:IDSrate}. We now apply Algo.~\ref{alg:ids} considering $r_{min} = 0.5$ for both VDC and TTC as derived from their respective settling time requirements. The value of $r_{min}$ combined with different possible values of $l$ provide us multiple  combinations of $(l, \theta)$ as given in Col. 2. 
For each case, Algo.~\ref{alg:ids} outputs the patterns with maximum resilience as provided in Col. 3 of Tab.~\ref{tab:IDSrate}. If there are multiple such  patterns with same resilience, the algorithm provides them in decreasing order of control performance (i.e. increasing LQR-D for the same $(l, \theta)$ in col. 6).  
For each pattern, the corresponding safe IDS configuration $\langle n_{up}, n_{down}\rangle$ is given in Col. 4 with the IDS execution rate in Col. 5.
\begin{wrapfigure}{l}{0.6\columnwidth}
\begin{subfigure}{0.65\columnwidth}
\includegraphics[width=\linewidth]{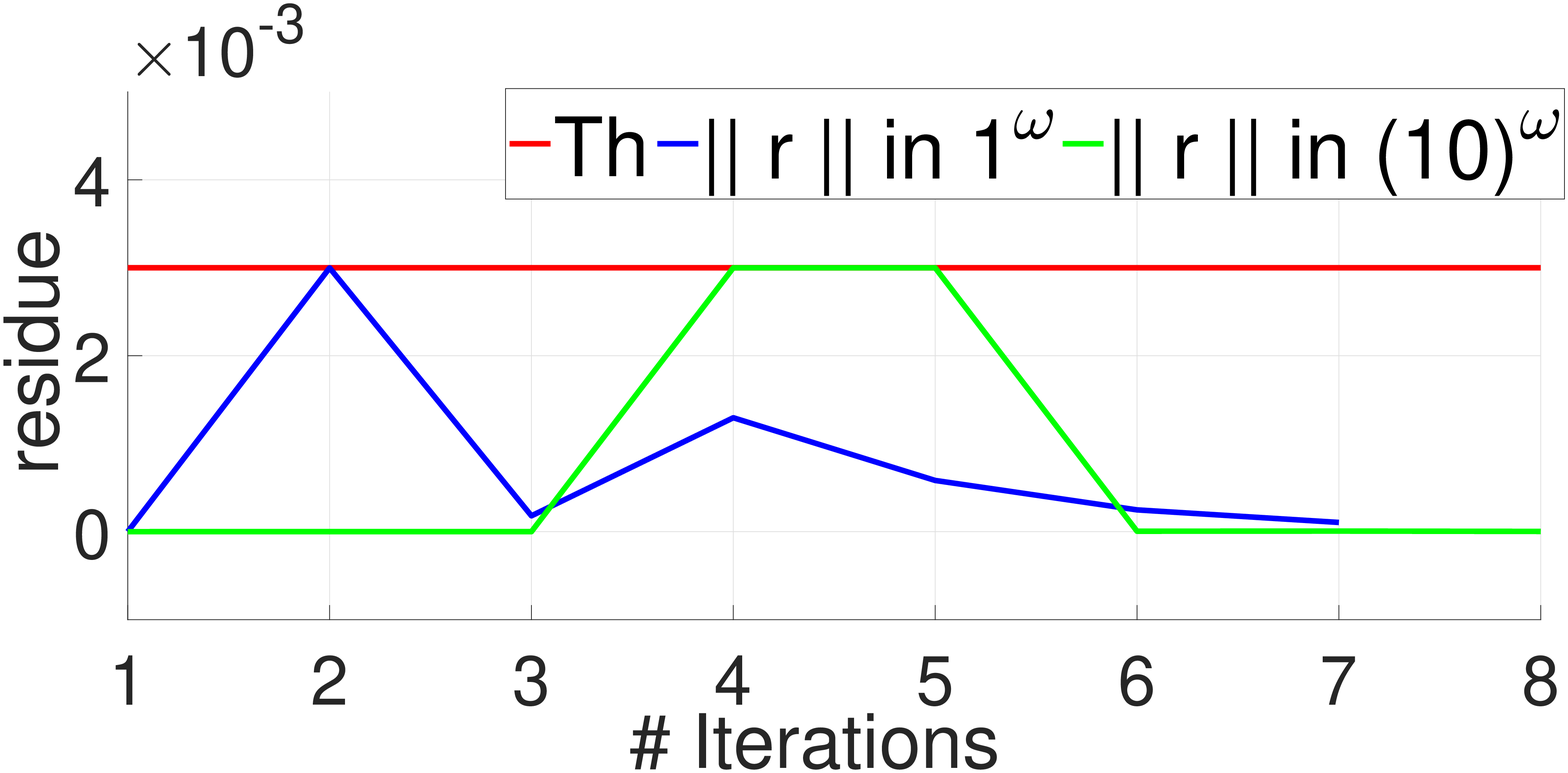}
    \caption{\footnotesize Stealthy attack on VDC}
    \label{fig:example_plot_r}
\end{subfigure}
\begin{subfigure}{0.7\columnwidth}
\includegraphics[width=\linewidth,clip]{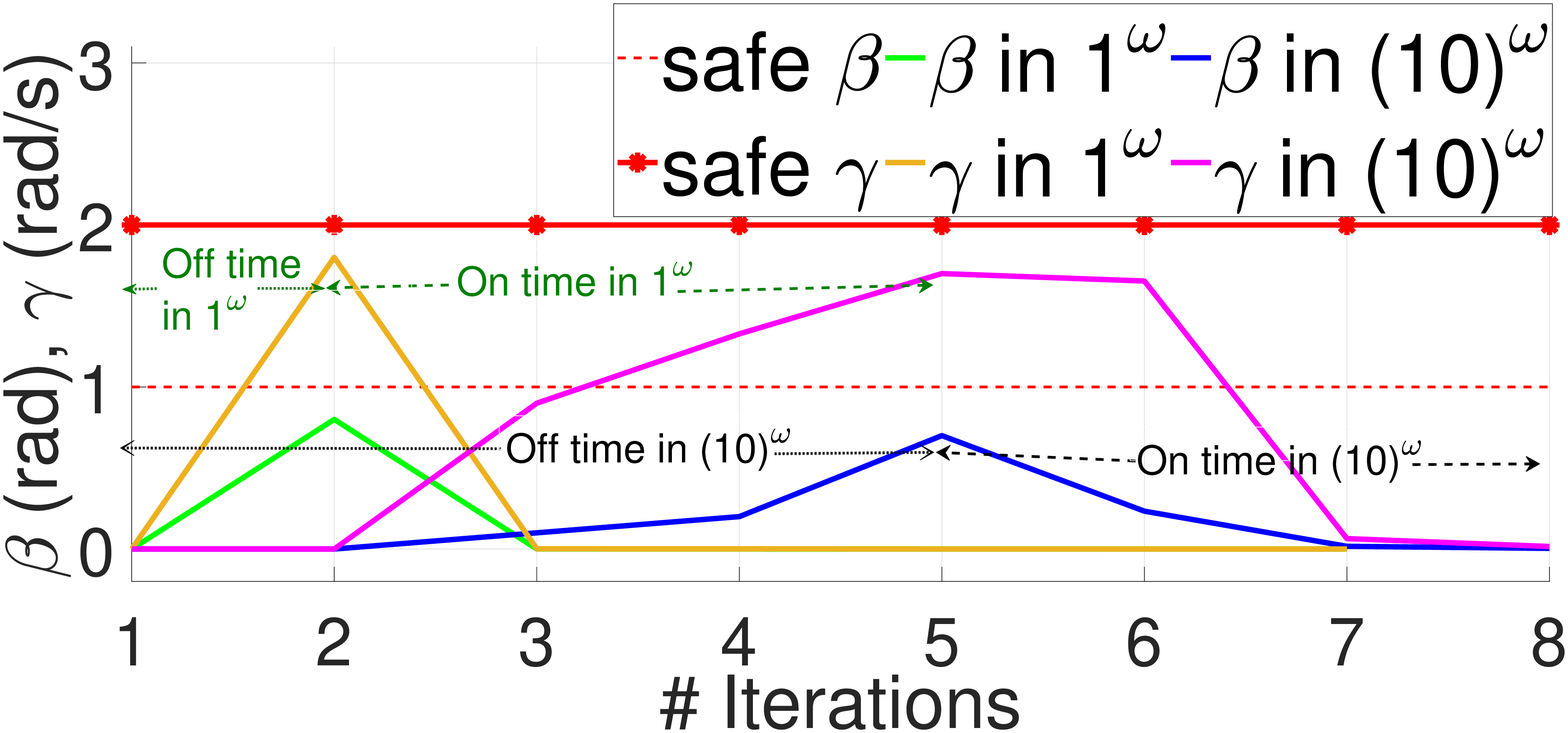}
    \caption{\footnotesize Higher IDS Off time ($n_{down}$) for control skipping}
    \label{fig:example_plot_state}
\end{subfigure}
\vspace{1.5pt}
\label{fig:example_plot_VDC}
\end{wrapfigure}
For each system, the patterns reported by our automated method as most attack resilient (i.e. requiring lowest IDS usage) are marked in bold. As  one may recall, the input $l$  and $r_{min}$ provides the maximum number of skips, i.e. $\theta_{max}$. For TTC, running our  method with $l=10$, we find  the most attack resilient pattern as $\rho=1010011111$ (with IDS rate $0.1667$)  showing a {$\mathbf{27.78\%}$} improvement w.r.t.  existing periodic IDS with $rate = 0.2307$ ( ref. Col. 5, Row 1). For $l=11$, we have $\rho=11010111100$ with similar resilience. 
For a given $l$, Algo.~\ref{alg:ids} (Lines \ref{alg:ids:foreachskip}-\ref{alg:ids:sortpbyskip}), automatically tries for different values of $\theta\in [1,\theta_{max}]$ and  reports only those values which provide better resilience w.r.t. periodic control. So, we do not have entries like $(l, \theta)=(10, 2)$  and many  others. Similarly for VDC, our methodology was tried with different values of $l$ and the most resilient solutions are shown in bold resulting in about $\mathbf{37.5\%}$ reduction in IDS rate.

\noindent For comparison, we consider the effect of a stealthy and successful attack on VDC when it is   executing the closed loop following $1^{\omega}$ (periodic) and $(10)^\omega$ (best pattern returned by Algo.~\ref{alg:ids} for $l=2$). Our method reveals the the minimum attack length for VDC following $1^{\omega}$ and $(10)^\omega$ as $3$ and $5$ respectively. Fig. \ref{fig:example_plot_r} shows the residue of the VDC  considering an attack scenario which is stealthy since $||r||\leq Th$ is always satisfied for both $1^\omega$ and $(10)^\omega$. For the same attack scenario, we plot system states (i.e., side slip $\beta$ and yaw rate $\gamma$) of the VDC in Fig. \ref{fig:example_plot_state} considering both $1^{\omega}$ and $(10)^\omega$. The attack inflicted during the IDS off time is unable to cross the safety limits (of value 1 and 2 in Y axis) as we activate IDS from $2$-nd iteration in case of $1^\omega$ and from $5$-th iteration in case of $(10)^\omega$ depending on their corresponding minimum attack lengths as mentioned earlier. The plot clearly demonstrates that due to the deployment of pattern based execution, the safety of the system is maintained in spite of increasing the down-time of the IDS (from $2$ to $5$). This validates our principal claim of potential increment in system attack resilience provably improving security by judiciously skipping some control executions.  Next, we demonstrate a useful application of the ability to implement provably safe sporadic IDS leveraging control skipping patterns in  automotive systems. 


\subsection{Manifestation on CAN bandwidth}
\label{subsec:Can_BW_util}
\begin{figure}[h]
\footnotesize
\centering
\includegraphics[width=\columnwidth,keepaspectratio,clip,scale=0.6]{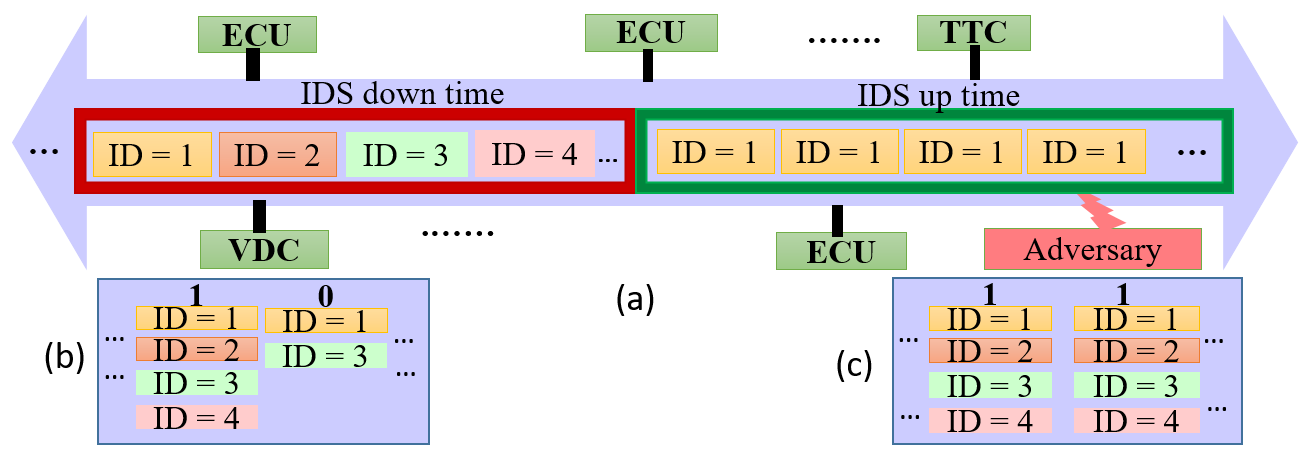}
\caption { \footnotesize \textbf{a)} CAN Transmissions with sporadic IDS in presence of adversary, \textbf{b)} Message flow for periodic execution, \textbf{c)} Message flow for skipped execution} \label{fig:can_example}
\vspace{1.2pt}
\end{figure}

Let us consider an automotive system where the CAN messages are communicated through the bus with a speed of $B$ bps at periodicity $p_1$, $p_2,\ldots,p_k$ such that  $p_1>p_2 >\cdots > p_k$. The  number of message types with rate $p_i$ is given by $m_i,\,i\in\{1,\cdots,k\}$. Assume that IDS is implemented for messages with periodicity $p_{k'}$ 
and there are $m_{k'} >0$   number of such types of messages. Similar to \cite{cookcontroller}, we consider a $p_1$-length observation window ($\geq$ the largest period) and compute  bandwidth consumption  in CAN bus for the aforementioned setup through the following steps.\\
~\textbf{A.} We find out the number of messages communicated over the observation window $p_1$. For any $m_i$ it is $c_i=\lceil p_1/p_i\rceil \forall i\in [0,k]$. We consider maximum CAN payload for each message, i.e. 64 bits.\\
~\textbf{B.} For each of the $m_k'$ different type of  messages, the IDS rate is $rate_i,i\in [1,m_k']$. If we design the IDS with CMAC/AES-128  (with $a$-bit CMAC) \cite{autosar_13} encryption  to provide confidentiality and authenticity, payload will be of size (64+$a$) bits. This will convert to $\lceil (64+$a$)/128\rceil$ AES blocks or $ $b$=(\lceil (64+$a$)/128\rceil\times 128)/64$  CAN frames (CAN payload size=64). In such an arrangement, each CAN frame will be replaced by $b$ CAN frames when IDS is active (refer Fig.~\ref{fig:can_example}a where $b=4$). Hence, over the observation window, each of the $m_{k'}$ messages is transmitted $(1-rate_i)\times c_{k'}$ times without IDS active and $b\times rate_i\times c_{k'}$ times with IDS active giving a total count of $(1+(b-1)rate_i)\times c_{k'}$.\\
~\textbf{C.} Additional $47$ bits are added to the payload to form one CAN frame (SOF + Arbitration + RTR + Control + CRC + Acknowledgment + EOF + Interframe Space = 1 + 11 + 1 + 6 + 16 + 2 + 7 + 3 = 47 bits)\cite{cookcontroller}. Thus, in our consideration, size of each CAN frame is (64+47) bits = 111 bits. Following this, total bandwidth consumption over \emph{observation window} is computed as $T= 111 \times[m_1+m_2\times c_2+..+\sum_{i=1}^{m_{k'}}(1+(b-1)rate_i)\times c_{k'}+..+m_k \times c_k]/B$.
Let the IDS rates for some control skipping pattern, output by Algo.~\ref{alg:ids} be  $rate'_i,\forall i\in[1,m_{k'}]$. Since Algorithm \ref{alg:ids} ensures if proposed patterns are used $rate'_i<rate_i(\forall i \in [1,m_{k'}])$, the improvement in bandwidth consumption when executing a pattern based schedule compared to a periodic schedule is given as, $\scriptsize(T-T')/T= 111\cdot \sum_{i=1}^{m_{k'}}((1+(b-1)(rate_i-rate'_i))\cdot c_3)/T $ considering $T'$ as the bandwidth consumed by pattern based schedule.
\par\noindent\textbf{\textit{Example:}} Let us consider the following setup of (\#message, periodicity): $\langle m_1,p_1\rangle=\langle 10,1\rangle,\langle m_2$,$p_2\rangle=\langle 20,0.2\rangle,\langle m_3,p_3\rangle=\langle 2,0.1\rangle(VDC),\\\langle m_4,p_4\rangle=\langle 2,0.1\rangle(TTC)$ in CAN bus. So, the VDC and TTC both require two types of messages (sensor o/p, control i/p) of period $p_3$ and $p_4$  respectively. These are denoted by CAN IDs $1\cdots 4$(Fig.~\ref{fig:can_example}a). 
During skips in the control execution, actuation signals are not communicated as we can see in Fig. 5c, which also frees bandwidth. 
If the IDS scheme in place uses $128$ bit CMAC (i.e. $a = 128$), it replaces each CAN frame with $b=4$ CAN frames when IDS is active (refer Fig. 5a). Following the derived formula for the aforementioned setup, we get \emph{16.25\% net  improvement in CAN bandwidth consumption} using the secure control schedule $10^\omega$ for VDC and $1010011111^\omega$ for TTC. Considering our methodology to design such pattern based secure control schedules for a significant number of control loops has an additive effect on the bandwidth saving. Thus our methodology helps to design sporadic IDS schemes  based on intentional control loop skips which promise better resource utilization in terms of communication bandwidth. 
\section{Related Work}
\label{sec:r_work}
We provide a brief survey on existing works in the area of secure control which are relevant in the context of the current work. In \cite{Mo1.1}, the authors discuss suitable   conditions under which a control system with $\chi^2$ based detectors is stealthily attackable. The performance degradation of such $\chi^2$ detector enabled systems in the presence of stealthy attacks has been quantified in \cite{sd2}. In \cite{sd3}, the authors report such `fake disturbance attacks' and their implications in Network Control Systems (NCS) in the presence of deterministic monitoring algorithms. The idea of stealthy attacks on both sensor and actuator sides being able to destabilize  automated power generation systems with threshold based detectors has been discussed in \cite{teixeira2015secure1}. Authors in \cite{carsten2015vehicle}
 also discuss security vulnerabilities in automotive CPS domain.
Designing resilient control implementations by leveraging secure state estimation techniques, more specifically in the automotive context has been reported in \cite{sd6}. The idea of sporadically using IDS schemes like MAC computation has been investigated in a different line of works \cite{jovanov2017sporadic, sd4, sd5}, but in the context of periodic control only. In \cite{ghosh2018performance}, the authors explore the advantage of employing lightweight periodic authentication schemes like Physically Unclonable Functions (PUFs) in the context of cyber physical security as a sporadically available IDS mechanism, again for periodic control. In that work, the periodic  availability of the authentication scheme depends on the PUF delay (PUF with high reliability incurs higher delay due to reliability peripherals like error correction, helper data etc). In the current work, we assume that the IDS security primitive is available for $n_{up}$ consecutive iterations followed by an off time for which we are able to establish a guarantee that the performance degradation due to stealthy attacks is inside recoverable limits.

\section{Conclusion}
\label{sec:conclu}
The present work 
demonstrates how control skipping patterns can be  synthesized guaranteeing desired performance with increased resilience. 
The safe and resilient patterns generated by the method helped in reducing the computation and communication overhead of 
IDS schemes employed in Automotive CPS. 
Integrating our SMT based technique with   safe but approximate analysis  (e.g. using `Barrier functions') can help increase the scalability of the approach for applicability in complex industrial test cases. This along with   controller synthesis for the joint objective of performance and security are important future extensions possible for this work. 

\bibliographystyle{ACM-Reference-Format}
\bibliography{Reference.bib}


\begin{thebibliography}{29}


\ifx \showCODEN    \undefined \def \showCODEN     #1{\unskip}     \fi
\ifx \showDOI      \undefined \def \showDOI       #1{#1}\fi
\ifx \showISBNx    \undefined \def \showISBNx     #1{\unskip}     \fi
\ifx \showISBNxiii \undefined \def \showISBNxiii  #1{\unskip}     \fi
\ifx \showISSN     \undefined \def \showISSN      #1{\unskip}     \fi
\ifx \showLCCN     \undefined \def \showLCCN      #1{\unskip}     \fi
\ifx \shownote     \undefined \def \shownote      #1{#1}          \fi
\ifx \showarticletitle \undefined \def \showarticletitle #1{#1}   \fi
\ifx \showURL      \undefined \def \showURL       {\relax}        \fi
\providecommand\bibfield[2]{#2}
\providecommand\bibinfo[2]{#2}
\providecommand\natexlab[1]{#1}
\providecommand\showeprint[2][]{arXiv:#2}

\bibitem[\protect\citeauthoryear{{\AA}str{\"o}m and Wittenmark}{{\AA}str{\"o}m
  and Wittenmark}{1997}]%
        {astrom97}
\bibfield{author}{\bibinfo{person}{Karl~J {\AA}str{\"o}m} {and}
  \bibinfo{person}{Bj{\"o}rn Wittenmark}.} \bibinfo{year}{1997}\natexlab{}.
\newblock \bibinfo{booktitle}{\emph{Computer-controlled systems}}.
\newblock \bibinfo{publisher}{Prentice-Hall, Inc.}
\newblock


\bibitem[\protect\citeauthoryear{Carsten and et~al.}{Carsten and
  et~al.}{2015}]%
        {carsten2015vehicle}
\bibfield{author}{\bibinfo{person}{Paul Carsten} {and} \bibinfo{person}{et
  al.}} \bibinfo{year}{2015}\natexlab{}.
\newblock \showarticletitle{In-vehicle networks: Attacks, vulnerabilities, and
  proposed solutions}. In \bibinfo{booktitle}{\emph{CISRC}}. ACM.
\newblock


\bibitem[\protect\citeauthoryear{Choffrut and Karhum{\"a}ki}{Choffrut and
  Karhum{\"a}ki}{1997}]%
        {choffrut1997combinatorics}
\bibfield{author}{\bibinfo{person}{Christian Choffrut} {and}
  \bibinfo{person}{Juhani Karhum{\"a}ki}.} \bibinfo{year}{1997}\natexlab{}.
\newblock \bibinfo{title}{Combinatorics of words, Handbook of formal languages,
  vol. 1: word, language, grammar}.
\newblock   (\bibinfo{year}{1997}).
\newblock


\bibitem[\protect\citeauthoryear{Cook and Freudenberg}{Cook and
  Freudenberg}{2007}]%
        {cookcontroller}
\bibfield{author}{\bibinfo{person}{JA Cook} {and} \bibinfo{person}{JS
  Freudenberg}.} \bibinfo{year}{2007}\natexlab{}.
\newblock \showarticletitle{Controller Area Network (CAN)}.
\newblock \bibinfo{journal}{\emph{EECS}}  \bibinfo{volume}{461}
  (\bibinfo{year}{2007}), \bibinfo{pages}{1--5}.
\newblock


\bibitem[\protect\citeauthoryear{De~Moura and Bj{\o}rner}{De~Moura and
  Bj{\o}rner}{2008}]%
        {de2008z3}
\bibfield{author}{\bibinfo{person}{Leonardo De~Moura} {and}
  \bibinfo{person}{Nikolaj Bj{\o}rner}.} \bibinfo{year}{2008}\natexlab{}.
\newblock \showarticletitle{Z3: An efficient SMT solver}. In
  \bibinfo{booktitle}{\emph{TACAS}}. Springer.
\newblock


\bibitem[\protect\citeauthoryear{Gerard et~al\mbox{.}}{Gerard
  et~al\mbox{.}}{2018}]%
        {sd2}
\bibfield{author}{\bibinfo{person}{Benjamin Gerard} {et~al\mbox{.}}}
  \bibinfo{year}{2018}\natexlab{}.
\newblock \showarticletitle{Cyber Security and Vulnerability Analysis of
  Networked Control System subject to False-Data injection}. In
  \bibinfo{booktitle}{\emph{ACC}}. IEEE.
\newblock


\bibitem[\protect\citeauthoryear{Ghosh et~al\mbox{.}}{Ghosh
  et~al\mbox{.}}{2017}]%
        {ghosh2017structured}
\bibfield{author}{\bibinfo{person}{Sumana Ghosh} {et~al\mbox{.}}}
  \bibinfo{year}{2017}\natexlab{}.
\newblock \showarticletitle{A structured methodology for pattern based adaptive
  scheduling in embedded control}.
\newblock \bibinfo{journal}{\emph{ACM TECS}} \bibinfo{volume}{16},
  \bibinfo{number}{5s} (\bibinfo{year}{2017}), \bibinfo{pages}{189}.
\newblock


\bibitem[\protect\citeauthoryear{Ghosh et~al\mbox{.}}{Ghosh
  et~al\mbox{.}}{2018}]%
        {ghosh2018performance}
\bibfield{author}{\bibinfo{person}{Saurav~K. Ghosh} {et~al\mbox{.}}}
  \bibinfo{year}{2018}\natexlab{}.
\newblock \showarticletitle{Performance, Security Trade-offs in Secure
  Control}.
\newblock \bibinfo{journal}{\emph{IEEE ESL}} (\bibinfo{year}{2018}).
\newblock


\bibitem[\protect\citeauthoryear{Giraldo et~al\mbox{.}}{Giraldo
  et~al\mbox{.}}{2018}]%
        {giraldo2018survey}
\bibfield{author}{\bibinfo{person}{Jairo Giraldo} {et~al\mbox{.}}}
  \bibinfo{year}{2018}\natexlab{}.
\newblock \showarticletitle{A survey of physics-based attack detection in
  cyber-physical systems}.
\newblock \bibinfo{journal}{\emph{ACM Computing Surveys (CSUR)}}
  \bibinfo{volume}{51}, \bibinfo{number}{4} (\bibinfo{year}{2018}),
  \bibinfo{pages}{76}.
\newblock


\bibitem[\protect\citeauthoryear{Jia, Song, and Simonot-Lion}{Jia
  et~al\mbox{.}}{2007}]%
        {jia2007graceful}
\bibfield{author}{\bibinfo{person}{Ning Jia}, \bibinfo{person}{Ye-Qiong Song},
  {and} \bibinfo{person}{Fran{\c{c}}oise Simonot-Lion}.}
  \bibinfo{year}{2007}\natexlab{}.
\newblock \showarticletitle{Graceful degradation of the quality of control
  through data drop policy}. In \bibinfo{booktitle}{\emph{2007 European Control
  Conference (ECC)}}. IEEE.
\newblock


\bibitem[\protect\citeauthoryear{Jovanov and Pajic}{Jovanov and Pajic}{2017}]%
        {jovanov2017sporadic}
\bibfield{author}{\bibinfo{person}{Ilija Jovanov} {and}
  \bibinfo{person}{Miroslav Pajic}.} \bibinfo{year}{2017}\natexlab{}.
\newblock \showarticletitle{Sporadic data integrity for secure state
  estimation}. In \bibinfo{booktitle}{\emph{CDC}}. IEEE.
\newblock


\bibitem[\protect\citeauthoryear{Jovanov and Pajic}{Jovanov and Pajic}{2018}]%
        {sd5}
\bibfield{author}{\bibinfo{person}{Ilija Jovanov} {and}
  \bibinfo{person}{Miroslav Pajic}.} \bibinfo{year}{2018}\natexlab{}.
\newblock \showarticletitle{Secure State Estimation with Cumulative Message
  Authentication}. In \bibinfo{booktitle}{\emph{CDC}}. IEEE.
\newblock


\bibitem[\protect\citeauthoryear{Kalman}{Kalman}{1960}]%
        {kalman1960new}
\bibfield{author}{\bibinfo{person}{Rudolph~Emil Kalman}.}
  \bibinfo{year}{1960}\natexlab{}.
\newblock \showarticletitle{A new approach to linear filtering and prediction
  problems}.
\newblock \bibinfo{journal}{\emph{J. Basic Eng.}} \bibinfo{volume}{82},
  \bibinfo{number}{1} (\bibinfo{year}{1960}), \bibinfo{pages}{35--45}.
\newblock


\bibitem[\protect\citeauthoryear{Koley et~al\mbox{.}}{Koley
  et~al\mbox{.}}{2020}]%
        {ipsita2020}
\bibfield{author}{\bibinfo{person}{Ipsita Koley} {et~al\mbox{.}}}
  \bibinfo{year}{2020}\natexlab{}.
\newblock \bibinfo{title}{Formal Synthesis of Monitoring and Detection Systems
  for Secure CPS Implementations}.
\newblock   (\bibinfo{year}{2020}).
\newblock
\showeprint[arxiv]{cs.CR/2002.12412}


\bibitem[\protect\citeauthoryear{Kreimel et~al\mbox{.}}{Kreimel
  et~al\mbox{.}}{2017}]%
        {kreimel2017anomaly}
\bibfield{author}{\bibinfo{person}{Philipp Kreimel} {et~al\mbox{.}}}
  \bibinfo{year}{2017}\natexlab{}.
\newblock \showarticletitle{Anomaly-Based Detection and Classification of
  Attacks in Cyber-Physical Systems}. In \bibinfo{booktitle}{\emph{ARES}}. ACM.
\newblock


\bibitem[\protect\citeauthoryear{Lesi et~al\mbox{.}}{Lesi
  et~al\mbox{.}}{2017}]%
        {sd4}
\bibfield{author}{\bibinfo{person}{Vuk Lesi} {et~al\mbox{.}}}
  \bibinfo{year}{2017}\natexlab{}.
\newblock \showarticletitle{Security-Aware Scheduling of Embedded Control
  Tasks}.
\newblock \bibinfo{journal}{\emph{ACM TECS}} \bibinfo{volume}{16},
  \bibinfo{number}{5} (\bibinfo{year}{2017}).
\newblock


\bibitem[\protect\citeauthoryear{Mo and Sinopoli}{Mo and Sinopoli}{2010}]%
        {Mo1.1}
\bibfield{author}{\bibinfo{person}{Yilin Mo} {and} \bibinfo{person}{Bruno
  Sinopoli}.} \bibinfo{year}{2010}\natexlab{}.
\newblock \showarticletitle{False data injection attacks in control systems}.
  In \bibinfo{booktitle}{\emph{SCS}}.
\newblock


\bibitem[\protect\citeauthoryear{Mo and Sinopoli}{Mo and Sinopoli}{2016}]%
        {sd3}
\bibfield{author}{\bibinfo{person}{Yilin Mo} {and} \bibinfo{person}{Bruno
  Sinopoli}.} \bibinfo{year}{2016}\natexlab{}.
\newblock \showarticletitle{On the Performance Degradation of Cyber-Physical
  Systems Under Stealthy Integrity Attacks}.
\newblock \bibinfo{journal}{\emph{IEEE TAC}} \bibinfo{volume}{61},
  \bibinfo{number}{9} (\bibinfo{year}{2016}), \bibinfo{pages}{2618--2624}.
\newblock


\bibitem[\protect\citeauthoryear{Motorsport}{Motorsport}{2020a}]%
        {yawrateDatasheet}
\bibfield{author}{\bibinfo{person}{Bosch Motorsport}.}
  \bibinfo{year}{2020}\natexlab{a}.
\newblock \bibinfo{title}{Acceleration Sensor MM5.10}.
\newblock   (\bibinfo{date}{May} \bibinfo{year}{2020}).
\newblock
\urldef\tempurl%
\url{http://www.bosch-motorsport.de/content/downloads/Raceparts/en-GB/51546379119226251.html}
\showURL{%
Retrieved May 28, 2020 from \tempurl}


\bibitem[\protect\citeauthoryear{Motorsport}{Motorsport}{2020b}]%
        {steeringAngleDatasheet}
\bibfield{author}{\bibinfo{person}{Bosch Motorsport}.}
  \bibinfo{year}{2020}\natexlab{b}.
\newblock \bibinfo{title}{Steering Wheel Angle Sensor LWS}.
\newblock   (\bibinfo{date}{May} \bibinfo{year}{2020}).
\newblock
\urldef\tempurl%
\url{http://www.bosch-motorsport.de/content/downloads/Raceparts/en-GB/54425995191962507.html}
\showURL{%
Retrieved May 28, 2020 from \tempurl}


\bibitem[\protect\citeauthoryear{Munir and Koushanfar}{Munir and
  Koushanfar}{2018}]%
        {munir2018design}
\bibfield{author}{\bibinfo{person}{Arslan Munir} {and} \bibinfo{person}{Farinaz
  Koushanfar}.} \bibinfo{year}{2018}\natexlab{}.
\newblock \showarticletitle{Design and analysis of secure and dependable
  automotive CPS: A steer-by-wire case study}.
\newblock \bibinfo{journal}{\emph{IEEE Transactions on Dependable and Secure
  Computing}} (\bibinfo{year}{2018}).
\newblock


\bibitem[\protect\citeauthoryear{Pajic et~al\mbox{.}}{Pajic
  et~al\mbox{.}}{2017}]%
        {sd6}
\bibfield{author}{\bibinfo{person}{Miroslav Pajic} {et~al\mbox{.}}}
  \bibinfo{year}{2017}\natexlab{}.
\newblock \showarticletitle{Design and Implementation of Attack-Resilient
  Cyberphysical Systems: With a Focus on Attack-Resilient State Estimators}.
\newblock \bibinfo{journal}{\emph{IEEE Control Systems Magazine}}
  \bibinfo{volume}{37}, \bibinfo{number}{2} (\bibinfo{date}{April}
  \bibinfo{year}{2017}), \bibinfo{pages}{66--81}.
\newblock


\bibitem[\protect\citeauthoryear{Soudbakhsh et~al\mbox{.}}{Soudbakhsh
  et~al\mbox{.}}{2013}]%
        {soudbakhsh_13}
\bibfield{author}{\bibinfo{person}{Damoon Soudbakhsh} {et~al\mbox{.}}}
  \bibinfo{year}{2013}\natexlab{}.
\newblock \showarticletitle{Co-design of control and platform with dropped
  signals}. In \bibinfo{booktitle}{\emph{ICCPS}}. ACM.
\newblock


\bibitem[\protect\citeauthoryear{Teixeira et~al\mbox{.}}{Teixeira
  et~al\mbox{.}}{2015a}]%
        {teixeira2015secure2}
\bibfield{author}{\bibinfo{person}{Andre Teixeira} {et~al\mbox{.}}}
  \bibinfo{year}{2015}\natexlab{a}.
\newblock \showarticletitle{A secure control framework for resource-limited
  adversaries}.
\newblock \bibinfo{journal}{\emph{Automatica}}  \bibinfo{volume}{51}
  (\bibinfo{year}{2015}), \bibinfo{pages}{135--148}.
\newblock


\bibitem[\protect\citeauthoryear{Teixeira et~al\mbox{.}}{Teixeira
  et~al\mbox{.}}{2015b}]%
        {teixeira2015secure1}
\bibfield{author}{\bibinfo{person}{Andre Teixeira} {et~al\mbox{.}}}
  \bibinfo{year}{2015}\natexlab{b}.
\newblock \showarticletitle{Secure control systems: A quantitative risk
  management approach}.
\newblock \bibinfo{journal}{\emph{IEEE Control Systems Magazine}}
  \bibinfo{volume}{35}, \bibinfo{number}{1} (\bibinfo{year}{2015}),
  \bibinfo{pages}{24--45}.
\newblock


\bibitem[\protect\citeauthoryear{Vatanparvar and Al~Faruque}{Vatanparvar and
  Al~Faruque}{2019}]%
        {vatanparvar2019self}
\bibfield{author}{\bibinfo{person}{Korosh Vatanparvar} {and}
  \bibinfo{person}{Mohammad~Abdullah Al~Faruque}.}
  \bibinfo{year}{2019}\natexlab{}.
\newblock \showarticletitle{Self-Secured Control with Anomaly Detection and
  Recovery in Automotive Cyber-Physical Systems}. In
  \bibinfo{booktitle}{\emph{DATE}}. IEEE.
\newblock


\bibitem[\protect\citeauthoryear{Wiesbaden}{Wiesbaden}{2013}]%
        {autosar_13}
\bibfield{author}{\bibinfo{person}{Springer~Fachmedien Wiesbaden}.}
  \bibinfo{year}{2013}\natexlab{}.
\newblock \showarticletitle{{AUTOSAR} --- The Worldwide Automotive Standard for
  E/E Systems}.
\newblock \bibinfo{journal}{\emph{ATZextra worldwide}} \bibinfo{volume}{18},
  \bibinfo{number}{9} (\bibinfo{date}{Oct} \bibinfo{year}{2013}),
  \bibinfo{pages}{5--12}.
\newblock


\bibitem[\protect\citeauthoryear{Zhang et~al\mbox{.}}{Zhang
  et~al\mbox{.}}{2001}]%
        {zhang01}
\bibfield{author}{\bibinfo{person}{Wei Zhang} {et~al\mbox{.}}}
  \bibinfo{year}{2001}\natexlab{}.
\newblock \showarticletitle{Stability of networked control systems}.
\newblock \bibinfo{journal}{\emph{IEEE Control Systems}} \bibinfo{volume}{21},
  \bibinfo{number}{1} (\bibinfo{date}{Feb} \bibinfo{year}{2001}),
  \bibinfo{pages}{84--99}.
\newblock


\bibitem[\protect\citeauthoryear{Zheng, Tang, Han, and Zhang}{Zheng
  et~al\mbox{.}}{2006}]%
        {zheng2006controller}
\bibfield{author}{\bibinfo{person}{Shuibo Zheng}, \bibinfo{person}{Houjun
  Tang}, \bibinfo{person}{Zhengzhi Han}, {and} \bibinfo{person}{Yong Zhang}.}
  \bibinfo{year}{2006}\natexlab{}.
\newblock \showarticletitle{Controller design for vehicle stability
  enhancement}.
\newblock \bibinfo{journal}{\emph{Control Engineering Practice}}
  \bibinfo{volume}{14}, \bibinfo{number}{12} (\bibinfo{year}{2006}),
  \bibinfo{pages}{1413--1421}.
\newblock


\end{thebibliography}
\end{document}